\documentclass[longauth]{aa} 

\usepackage{graphicx}
\usepackage{txfonts}
\usepackage[utf8]{inputenc}
\usepackage{array}
\usepackage{float}
\usepackage{multirow}
\usepackage[]{natbib}
\usepackage{amsmath}
\usepackage{epstopdf}
\usepackage[colorlinks=true,linkcolor=blue,citecolor=blue]{hyperref}
\begin{document} 

\title{Hint {of} curvature in the orbital motion of the exoplanet 51~Eridani~b using 3 years of VLT/SPHERE monitoring\thanks{Based on observations collected at the European Organisation for Astronomical Research in the Southern Hemisphere under ESO programmes 095.C-0298, 096.C-0241, 198.C-0209, and 1100.C-0481.}}

  \author{A.-L. Maire\inst{1,2,\thanks{F.R.S.-FNRS Postdoctoral Researcher.}}, L. Rodet\inst{3}, F. Cantalloube\inst{1}, R. Galicher\inst{4}, W. Brandner\inst{1}, S. Messina\inst{5}, C. Lazzoni\inst{6,7}, D. Mesa\inst{6,8}, D. Melnick\inst{9}, J. Carson\inst{9,1}, M. Samland\inst{1,\thanks{International Max Planck Research School for Astronomy and Cosmic Physics, Heidelberg, Germany}}, B.~A. Biller\inst{10,1}, A. Boccaletti\inst{4}, Z. Wahhaj\inst{11,12}, H. Beust\inst{3}, M. Bonnefoy\inst{3}, G. Chauvin\inst{3,13}, S. Desidera\inst{6}, M. Langlois\inst{14,12}, T. Henning\inst{1}, M. Janson\inst{15,1}, J. Olofsson\inst{16,17,1}, D. Rouan\inst{4}, F. M\'enard\inst{3}, A.-M. Lagrange\inst{3}, R. Gratton\inst{6}, A. Vigan\inst{12}, M.~R. Meyer\inst{18,19}, A. Cheetham\inst{20}, J.-L. Beuzit\inst{12,3}, K. Dohlen\inst{12}, H. Avenhaus\inst{1,19}, M. Bonavita\inst{10,6}, R. Claudi\inst{6}, M. Cudel\inst{3}, S. Daemgen\inst{19}, V. D'Orazi\inst{6}, C. Fontanive\inst{10}, J. Hagelberg\inst{3,19}, H. Le Coroller\inst{12}, C. Perrot\inst{4,16,17}, E. Rickman\inst{20}, T. Schmidt\inst{4,21}, E. Sissa\inst{6}, S. Udry\inst{20}, A. Zurlo\inst{22,23,12,6}, L. Abe\inst{24}, A. Orign\'e\inst{12}, F. Rigal\inst{25}, G. Rousset\inst{4}, A. Roux\inst{3}, and L. Weber\inst{20}
         }

   \institute{Max-Planck-Institut f\"ur Astronomie, K\"onigstuhl 17, D-69117 Heidelberg, Germany
          \and
          STAR Institute, Universit\'e de Li\`ege, All\'ee du Six Ao\^ut 19c, B-4000 Li\`ege, Belgium \\
          \email{almaire@uliege.be}
         \and
         Univ. Grenoble Alpes, CNRS, IPAG, F-38000 Grenoble, France
         \and
         LESIA, Observatoire de Paris, Universit\'e PSL, CNRS, Sorbonne Universit\'e, Univ. Paris Diderot, Sorbonne Paris Cit\'e, 5 place Jules Janssen, F-92195 Meudon, France
         \and
         INAF Catania Astrophysical Observatory, via S. Sofia 78, 95123 Catania, Italy
         \and
         INAF--Osservatorio Astronomico di Padova, Vicolo dell'Osservatorio 5, I-35122 Padova, Italy
         \and
         Dipartimento di Fisica e Astronomia ``G. Galilei'', Universit\`a di Padova, Via Marzolo, 8, 35121 Padova, Italy
         \and
         INCT, Universidad de Atacama, calle Copayapu 485, Copiap\'o, Atacama, Chile
         \and
         Department of Physics \& Astronomy, College of Charleston, 66 George Street, Charleston, SC 29424, USA
         \and
         Institute for Astronomy, The University of Edinburgh, Royal Observatory, Blackford Hill View, Edinburgh, EH9 3HJ, UK
         \and
         European Southern Observatory, Alonso de Cordova 3107, Casilla 19001 Vitacura, Santiago 19, Chile
         \and
         Aix Marseille Universit\'e, CNRS, CNES,  LAM, Marseille, France
         \and
         Unidad Mixta Internacional Franco-Chilena de Astronom\'ia CNRS/INSU UMI 3386 and Departamento de Astronom\'ia, Universidad de Chile, Casilla 36-D, Santiago, Chile
         \and
         CRAL, UMR 5574, CNRS/ENS-Lyon/Universit\'e Lyon 1, 9 av. Ch. Andr\'e, F-69561 Saint-Genis-Laval, France
                \and
         Department of Astronomy, Stockholm University, AlbaNova University Center, 106 91 Stockholm, Sweden
                \and
                Instituto de F\'isica y Astronom\'ia, Facultad de Ciencias, Universidad de Valpara\'iso, Av. Gran Breta\~{n}a 1111, Playa Ancha, Valpara\'iso, Chile
                \and
                N\'ucleo Milenio Formaci\'on Planetaria - NPF, Universidad de Valpara\'iso, Av. Gran Breta\~{n}a 1111, Playa Ancha, Valpara\'iso, Chile
                \and
                Department of Astronomy, University of Michigan, 1085 S. University Ave, Ann Arbor, MI 48109-1107, USA
                \and
                Institute for Particle Physics and Astrophysics, ETH Zurich, Wolfgang-Pauli-Strasse 27, 8093 Zurich, Switzerland
                \and
         Geneva Observatory, University of Geneva, Chemin des Maillettes 51, 1290 Versoix, Switzerland
                \and
                Hamburger Sternwarte, Gojenbergsweg 112, 21029 Hamburg, Germany
         \and
         N\'ucleo de Astronom\'ia, Facultad de Ingenier\'ia y Ciencias, Universidad Diego Portales, Av. Ejercito 441, Santiago, Chile
         \and
         Escuela de Ingenier\'ia Industrial, Facultad de Ingenier\'ia y Ciencias, Universidad Diego Portales, Av. Ejercito 441, Santiago, Chile
         \and
         Universit\'e C\^ote d'Azur, OCA, CNRS, Lagrange, France
         \and
         NOVA Optical Infrared Instrumentation Group, Oude Hoogeveensedijk 4, 7991 PD Dwingeloo, The Netherlands
            }

   \date{Received 7 January 2019 / Accepted 15 March 2019}

 
  \abstract
   {The 51~Eridani system harbors a complex architecture with its primary star forming a hierarchical system with the binary GJ~3305AB at a projected separation of 2000~au, a giant planet orbiting the primary star at 13~au, and a low-mass debris disk around the primary star with possible cold and warm components inferred from the spectral energy distribution.}
   {We aim to better constrain the orbital parameters of the known giant planet.}
   {We monitored the system over three years from 2015 to 2018 with the {Spectro-Polarimetric High-contrast Exoplanet REsearch (SPHERE) instrument at the Very Large Telescope (VLT)}.}
  {We measure an orbital motion for the planet of $\sim$130~mas with a slightly decreasing separation ($\sim$10~mas) and find a {hint of} curvature. This potential curvature is further supported {at 3$\sigma$ significance when including literature {Gemini Planet Imager (GPI)} astrometry corrected for} calibration systematics. Fits of the SPHERE and GPI data using three complementary approaches provide {broadly similar results}. The data suggest an orbital period of {32$^{+17}_{-9}$~yr} (i.e., {12$^{+4}_{-2}$~au} in semi-major axis), an inclination of {133$^{+14}_{-7}$~deg}, an eccentricity of {0.45$^{+0.10}_{-0.15}$}, and an argument of periastron passage of {87$^{+34}_{-30}$~deg} [mod 180$^{\circ}$]. {The time at periastron passage and the longitude of node exhibit bimodal distributions} because we do not yet detect whether the planet is accelerating or decelerating {along its orbit. Given the inclinations of the orbit and of the stellar rotation axis ({134--144$^{\circ}$}), we infer alignment or misalignment within 18$^{\circ}$ for the star--planet spin-orbit}. Further astrometric monitoring in the next 3--4 years is required to confirm at a higher significance the curvature in the motion of the planet, determine if the planet is accelerating or decelerating on its orbit, and further constrain its orbital parameters {and the star--planet spin-orbit}.}
   {}


\authorrunning{A.-L. Maire et al.}
\titlerunning{VLT/SPHERE astrometric monitoring of the giant exoplanet 51 Eridani b}

   \maketitle

\section{Introduction}

\begin{table*}[t]
\caption{Observing log of SPHERE observations of 51~Eridani.}
\label{tab:obs}
\begin{center}
\begin{tabular}{l c c c c c c c c}
\hline\hline
UT date & $\epsilon$ ($''$) & $\tau_0$ (ms) & AM start/end & Mode & Bands & DIT\,(s)\,$\times$\,Nfr & FoV rot. ($^{\circ}$) & SR \\
\hline
2015/09/25 & 0.5--1.0 & 4--9 & 1.10--1.09 & IRDIFS\_EXT & $YJH$+$K12$ & 16$\times$256 & 41.5 & 0.68--0.88 \\
2015/09/26 & 0.7--1.3 & 6--12 & 1.10--1.09 & IRDIFS & $YJ$+$BBH$ & 4(64)$\times$918(64) & 42.6 & 0.66-0.90 \\
2016/01/16 & 1.6--2.3 & 1 & 1.09--1.10 & IRDIFS & $YJ$+$H23$ & 16(64)$\times$256(64) & 41.8 & 0.63--0.86 \\
2016/12/12 & 1.6--2.8 & 2 & 1.09--1.16 & IRDIFS & $YJ$+$H23$ & 64$\times$54 & 25.2 & 0.55--0.62 \\
2016/12/13 & 0.6--1.0 & 4--8 & 1.10--1.09 & IRDIFS & $YJ$+$H23$ & 64$\times$72 & 44.4 & 0.78--0.92 \\
2017/09/28 & 0.4--0.7 & 5--12 & 1.10--1.09 & IRDIFS\_EXT & $YJH$+$K12$ & 24(32)$\times$192(144) & 44.1 & 0.85--0.91 \\
2018/09/18 & 0.7--1.2 & 2--5 & 1.21--1.08 & IRDIFS\_EXT & $YJH$+$K12$ & 24(32)$\times$200(160) & 38.5 & 0.64--0.87 \\
\hline
\end{tabular}
\end{center}
\tablefoot{The columns provide the observing date, the seeing and coherence time measured by the differential image motion monitor (DIMM) at 0.5~$\muup$m, the airmass at the beginning and the end of the sequence, the observing mode, the spectral bands, the DIT (detector integration time) multiplied by the number of frames in the sequence, the field of view rotation, and the Strehl ratio measured by the adaptive optics system at 1.6~$\muup$m. For the DIT$\times$Nfr column, the numbers {in parentheses} are for the IFS data.}
\end{table*} 

The giant planet 51~Eridani~b is the first of its kind discovered in the {Gemini Planet Imager (GPI)} exoplanet imaging survey \citep{Macintosh2015}. {The methane-rich planet is a bound companion to the young star \object{51~Eridani}, which is a member of the 24-Myr $\beta$ Pictoris moving group \citep{Zuckerman2001, Torres2008, Bell2015}}. {The star is located} at 29.78$\pm$0.15~pc{, inferred from the inverse of the parallax measured by \textit{Gaia} \citep{GaiaCollaboration2018}. It is in good agreement with the value derived with an optimized approach \citep[29.76$\pm$0.12~pc,][]{BailerJones2018}. {Our uncertainty of 0.15~pc includes in addition to the statistical error of 0.12~pc an uncertainty term of 0.1~mas to account for potential parallax systematics (\url{https://www.cosmos.esa.int/web/gaia/dr2}).} The star} forms a hierarchical system with the M-dwarf binary \object{GJ~3305}AB with separation $\sim$10~au located at a projected separation of 2000~au {\citep{Feigelson2006, Montet2015}}. {\citet{Simon2011} measured a stellar radius of 1.63$\pm$0.03~$R_{\odot}$ with the {Center for High Angular Resolution Astronomy (CHARA)} interferometer and inferred a stellar mass of 1.75$\pm$0.05~$M_{\odot}$.} The primary star also harbors a debris disk inferred from the spectral energy distribution \citep[{SED;}][]{RiviereMarichalar2014, Patel2014}. \citet{RiviereMarichalar2014} estimated a low infrared (IR) fractional luminosity $L_{\rm{IR}}$/$L_{\odot}$\,=\,2.3$\times$10$^{-6}$ from \textit{Herschel} photometry. Since their analysis is based on fitting a three-parameter model of a modified blackbody to three data points with excess IR emission at wavelengths $\ge$70~$\muup$m, the resulting value for the inner edge of the cold dust belt is largely uncertain with 82$^{+677}_{-75}$~au. They also estimated an upper limit for the dust mass of 1.6$\times$10$^{-3}$~$M_{\oplus}$ and did not report gas detection ([OI], [CII]). \citet{Patel2014} observed the target with WISE as part of a survey for warm debris disks and inferred a warm disk with a temperature of 180~K (upper limit 344~K) and a radius of 5.5~au (lower limit 1.5~au) assuming the disk radiates as a blackbody. The {51~Eridani system} could therefore harbor a two-belt debris disk architecture, a feature observed in other young systems with giant planets, such as HR\,8799 \citep{Marois2008c, Marois2010b, Su2009} and HD\,95086 \citep{Rameau2013c, Moor2013}.

The planet 51~Eridani~b has a projected separation of $\sim$13~au from the primary star. \citet{Macintosh2015} could not confirm the companionship with a proper motion test because of the very short time baseline of their GPI measurements ($\sim$1.5 months, between December 2014 and January 2015). Instead, the planetary nature hypothesis is based on the spectrum {showing methane absorption}. \citet{DeRosa2015} presented a new GPI astrometric epoch obtained in September 2015, confirming that the planet is gravitationally bound, and detected orbital motion. They also carried out a preliminary assessment of its orbital elements using Bayesian rejection sampling and Markov-chain Monte Carlo methods. Their analysis suggests {most probable values with 1$\sigma$ error bars} for the semi-major axis of 14$^{+7}_{-3}$~au, for the period of 41$^{+35}_{-12}$~yr, and for the inclination of 138$^{+15}_{-13}$~deg. The other parameters are marginally constrained. The authors also noted that the orbital inclination of the planet is different from the inclination of the orbital plane of the binary GJ~3305AB \citep[$i$\,=\,92.1$\pm$0.2$^{\circ}$,][]{Montet2015}, implying that they cannot be coplanar.

We present in this paper astrometric follow-up observations of 51~Eridani~b obtained with the {Spectro-Polarimetric High-contrast Exoplanet REsearch (SPHERE) instrument {\citep{Beuzit2019}} at the Very Large Telescope (VLT)} as part of the SpHere INfrared survey for Exoplanets \citep[SHINE;][]{Chauvin2017b}. We describe the observations and the data reduction (Sect.~\ref{sec:data}).  We then use the new astrometric data of the planet to analyze its orbital motion (Sect.~\ref{sec:astrometry}). We subsequently fit the SPHERE astrometry in combination with GPI data to derive its orbital parameters (Sect.~\ref{sec:orbit}).

\begin{table*}[t]
\caption{SPHERE astrometry relative to the star of 51~Eridani~b.}
\label{tab:astrometry}
\begin{center}
\begin{tabular}{l c c c c c c c}
\hline\hline
Epoch & Filter & $\rho$ & PA & $\Delta$RA & $\Delta$Dec & Pixel scale & North correction angle \\
 & & (mas) & ($^{\circ}$) & (mas) & (mas) & (mas/pix) & ($^{\circ}$) \\
\hline
2015.74 & $K1$ & 453.4$\pm$4.6 & 167.15$\pm$0.56 & 100.8$\pm$2.9 & $-$442.0$\pm$3.6 & 12.267$\pm$0.009 & $-$1.813$\pm$0.046\\
2015.74 & $H$ & 453.9$\pm$16.3 & 166.1$\pm$2.0 & 108.7$\pm$8.8 & $-$440.7$\pm$13.7 & 12.251$\pm$0.009 & $-$1.813$\pm$0.046\\
{2016.04} & $H2$ & 456.7$\pm$6.9 & 165.50$\pm$0.84 & 114.3$\pm$4.5 & $-$442.2$\pm$5.2 & 12.255$\pm$0.009 & $-$1.82$\pm$0.06\\
2016.95 & $H2$ & 453.6$\pm$5.7 & 160.30$\pm$0.72 & 152.9$\pm$3.4 & $-$427.1$\pm$4.6 & 12.255$\pm$0.009 & $-$1.808$\pm$0.043\\
2017.74 & $K1$ & 449.0$\pm$2.9 & 155.67$\pm$0.38 & 185.0$\pm$2.0 & $-$409.2$\pm$2.1 & 12.267$\pm$0.009 & $-$1.735$\pm$0.043\\
2018.72 & $K1$ & 443.3$\pm$4.2 & 150.23$\pm$0.55 & 220.2$\pm$2.8 & $-$384.8$\pm$3.1 & 12.267$\pm$0.009 & $-$1.796$\pm$0.068\\
\hline
\end{tabular}
\end{center}
\tablefoot{The astrometric error bars were derived assuming an error budget including the measurement uncertainties (image post-processing) and the systematic uncertainties (calibration). The uncertainties in the estimation of the star location for the sequences obtained {without the stellar replicas in the science images} were estimated using calibration data taken before and after the science images (see text). The values are 0.32, 7.87, and 2.02~mas for the 2015 September 25, 2015 September 26, and January 2016 datasets, respectively. \\
}
\end{table*}

\section{Observations and data analysis}
\label{sec:data}

We observed 51~Eridani eight times from September 2015 to September 2018 with the IRDIFS mode of SPHERE. In this mode, the near-infrared (NIR) camera IRDIS \citep{Dohlen2008a, Vigan2010} and integral field spectrograph IFS \citep{Claudi2008} are operated in parallel, either in the $YJ$ bands for IFS and the $H23$ filter pair for IRDIS (standard IRDIFS mode) or in the $YJH$ bands for IFS and the $K12$ filter pair for IRDIS (IRDIFS\_EXT mode). Four datasets were published in an analysis of the SED of the planet in \citet{Samland2017}. Table~\ref{tab:obs} lists the published observations used for astrometry and the new observations. {We only considered the IRDIS data in this work because the planet astrometry could be extracted from a higher number of datasets due to limitations in signal-to-noise ratio}. The challenging contrast of the planet meant that it could be detected and its astrometry measured in six datasets only (Table~\ref{tab:astrometry}).

For all sequences, an apodized pupil Lyot coronagraph \citep{Carbillet2011, Martinez2009} was used. For calibrating the flux and the centering of the images, we acquired unsaturated non-coronagraphic images of the star (hereafter reference point-spread function or reference PSF) and coronagraphic images with four artificial crosswise replicas of the star \citep{Langlois2013} at the beginning and end of the sequences. For all datasets obtained starting from December 2016, the science images were recorded with the stellar replicas simultaneously to minimize the frame centering uncertainties in the astrometric error budget. {Night-time sky background frames were taken and additional daytime calibration performed following the standard procedure at ESO.}

\begin{figure}[t]
\centering
\includegraphics[width=.22\textwidth]{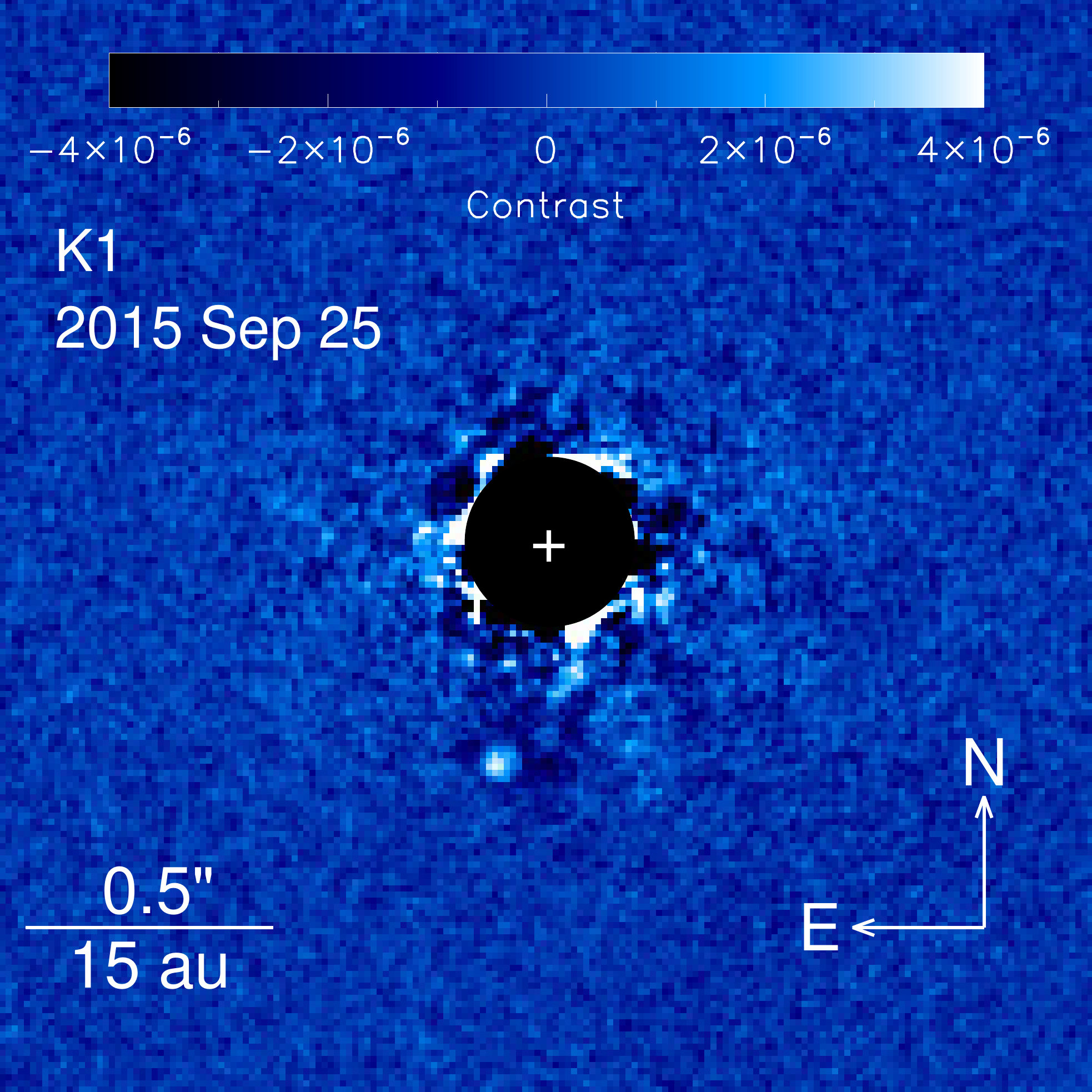}
\includegraphics[width=.22\textwidth]{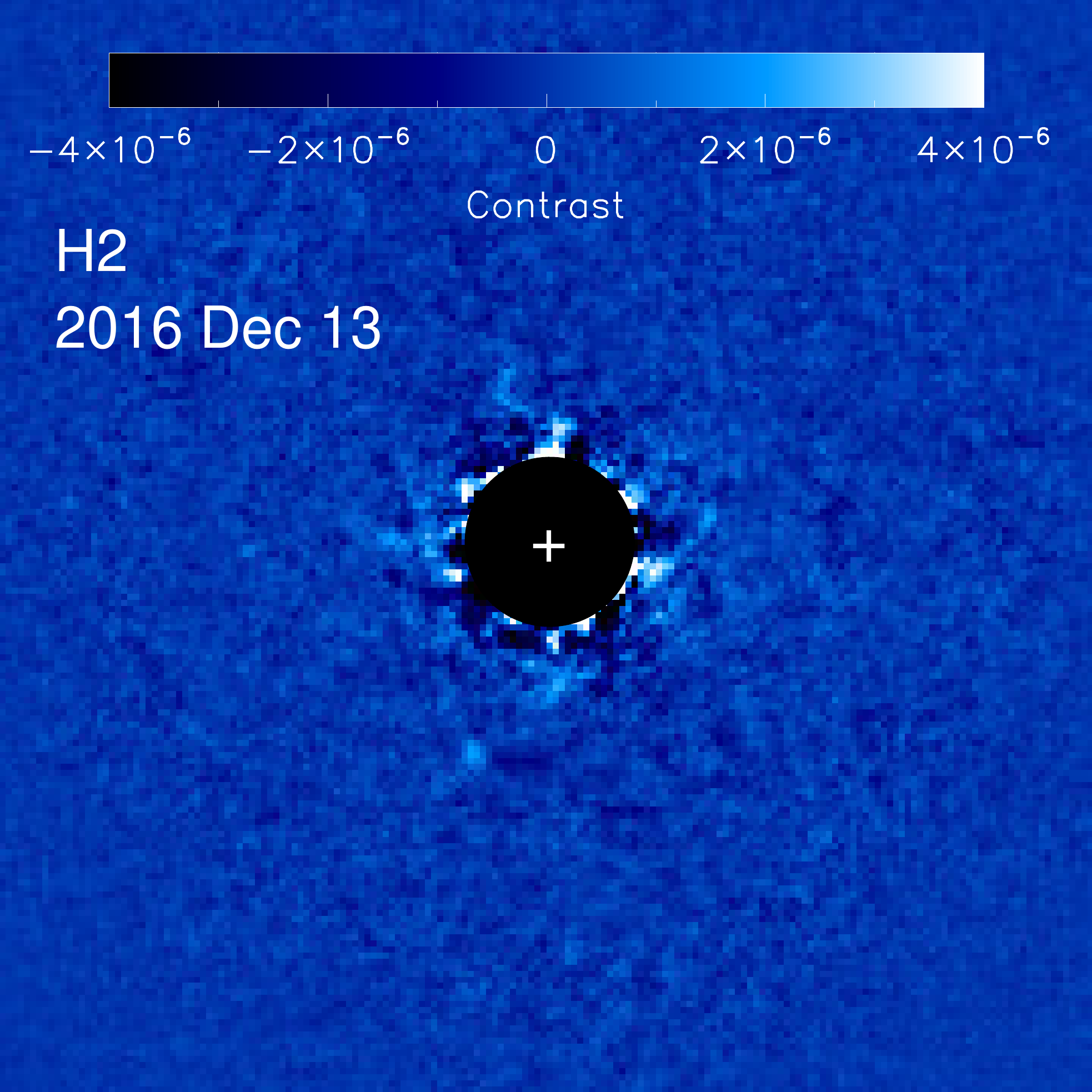}
\includegraphics[width=.22\textwidth]{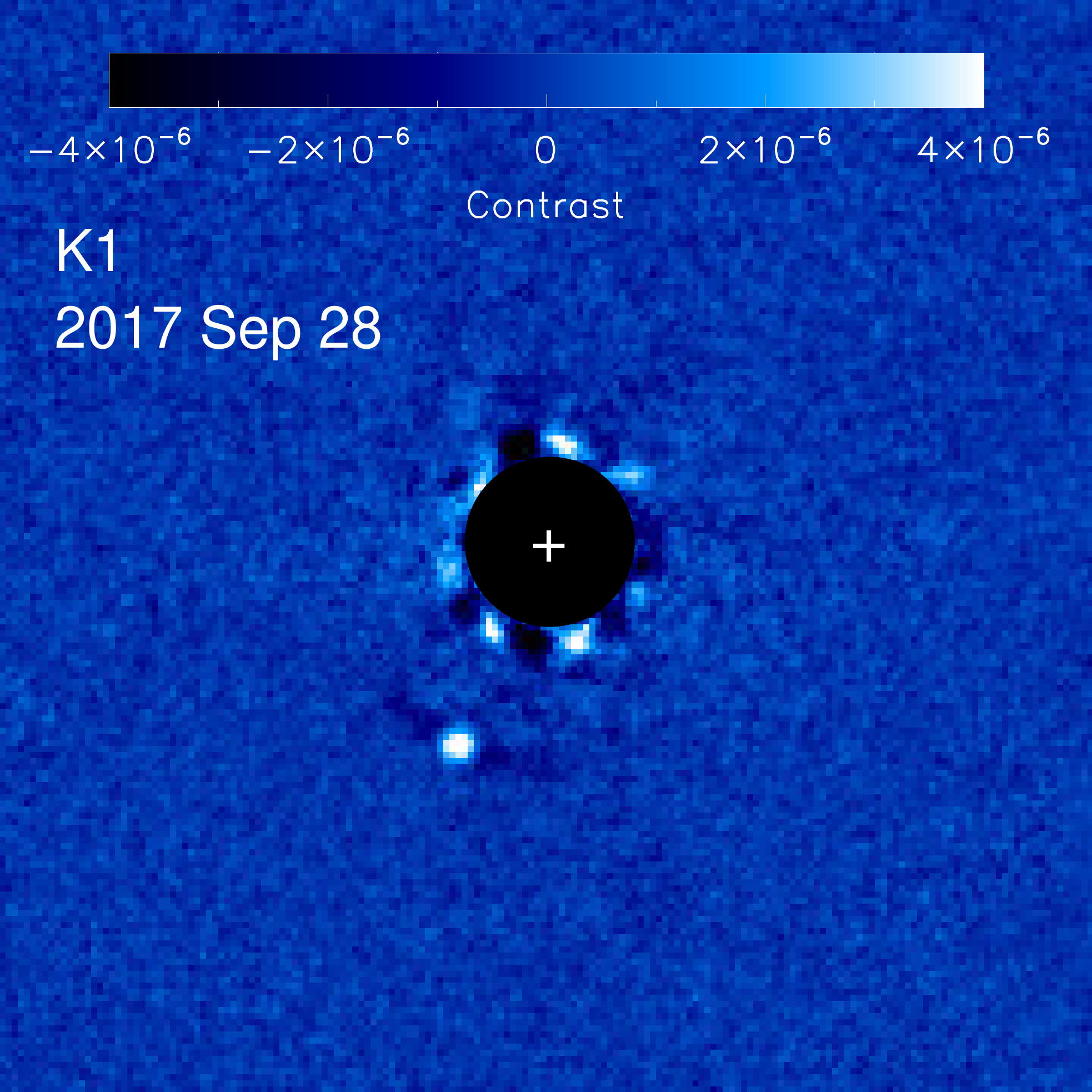}
\includegraphics[width=.22\textwidth]{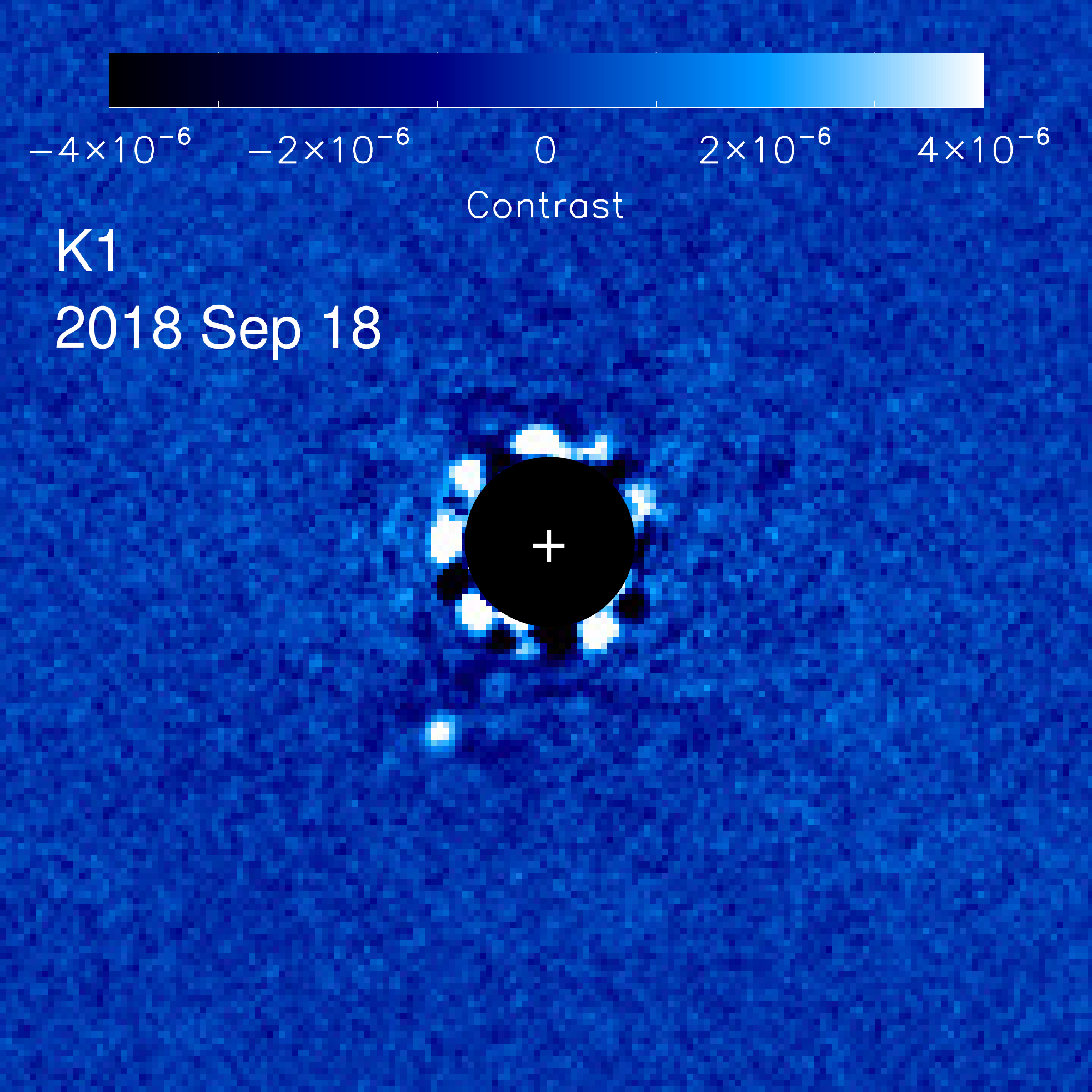}
\caption{SPHERE/IRDIS SDI+TLOCI contrast images of 51~Eridani at four epochs obtained with a narrow-band filter in $H2$ ($\lambda_{H2}$\,=\,1.593~$\muup$m, December 2016) and in $K1$ ($\lambda_{K1}$\,=\,2.110~$\muup$m, all other epochs). The central regions of the images were numerically masked out to hide bright stellar residuals. The white crosses indicate the location of the star.}
\label{fig:51eri_images}
\end{figure}

The data were reduced with the SPHERE Data Center pipeline \citep{Delorme2017b}, which uses the Data Reduction and Handling software \citep[v0.15.0,][]{Pavlov2008} and custom routines. It corrects for the cosmetics and instrument distortion, registers the frames, and normalizes their flux. Subsequently, we sorted the frames using visual inspection to reject poor-quality frames (adaptive optics open loops, low-wind effect) and an automatic criterion to reject frames with low flux in the coronagraphic spot (semi-transparent mask). After this step, we were left with 77--97\% of the frames, depending on the sequence. Finally, the data were analyzed with a consortium image processing pipeline \citep{Galicher2018}. Figure~\ref{fig:51eri_images} shows the IRDIS images obtained for the best epochs with a two-step process\footnote{The 2015 September 26 dataset was obtained with the broad $H$-band filter, so it was processed with angular differential imaging only.}: simultaneous spectral differential imaging \citep[SDI;][]{Racine1999} and angular differential imaging \citep{Marois2006a} with the Template Locally Optimized Combination of Images algorithm \citep[TLOCI;][]{Marois2014}.

\begin{figure}[t]
\centering
\includegraphics[width=.49\textwidth]{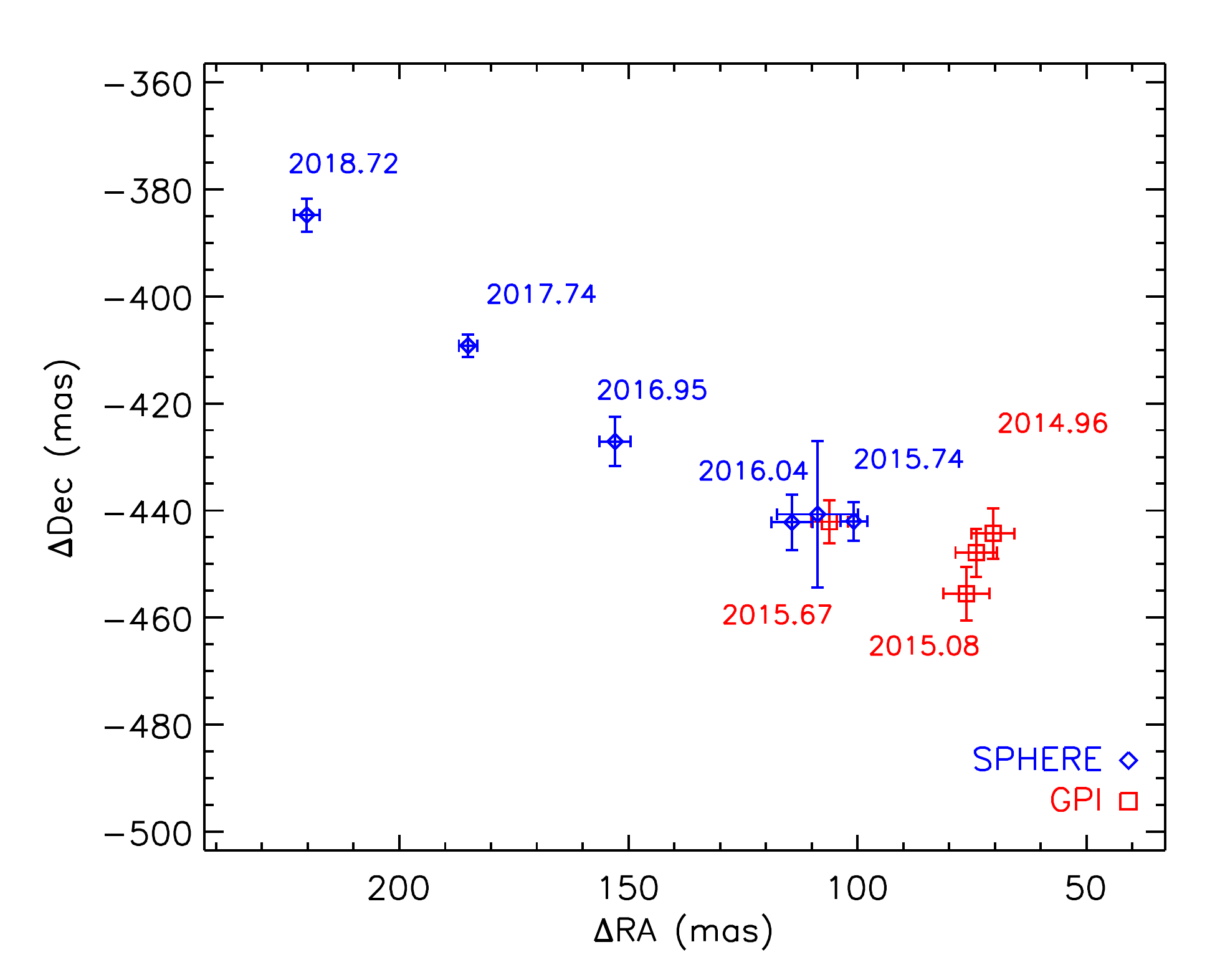}
\caption{Compilation of the astrometric measurements of 51~Eridani~b. {The GPI data are taken from \citet{DeRosa2015} without recalibration on the SPHERE data.}}
\label{fig:51eri_astrometry}
\end{figure}

{For all epochs, the planet astrometry and photometry was measured in the SDI+TLOCI images using the fit of a model of planet image built from the reference PSF and processed with TLOCI \citep{Galicher2018}. The position and flux of the model of planet image was optimized to minimize the image residuals within a circular region of radius 1.5 full width at half maximum centered on the measured planet location.} The values reported in Table~\ref{tab:astrometry} were calibrated following the methods in \citet{Maire2016b}. We also compared them with the astrometry extracted using SDI in combination with the ANgular DiffeRential Optimal Method Exoplanet Detection Algorithm \citep[ANDROMEDA,][]{Mugnier2009, Cantalloube2015} and found most values to agree within the TLOCI measurement uncertainties (Appendix~\ref{sec:otheralgos}). {We use the SDI+TLOCI astrometry for the astrometric and orbital analyses in the following sections, because TLOCI was tested and validated on a larger number of SPHERE datasets to retrieve the astrometry and photometry of detected companions \citep{Galicher2018}}.

\begin{figure*}[t]
\centering
\includegraphics[width=.42\textwidth]{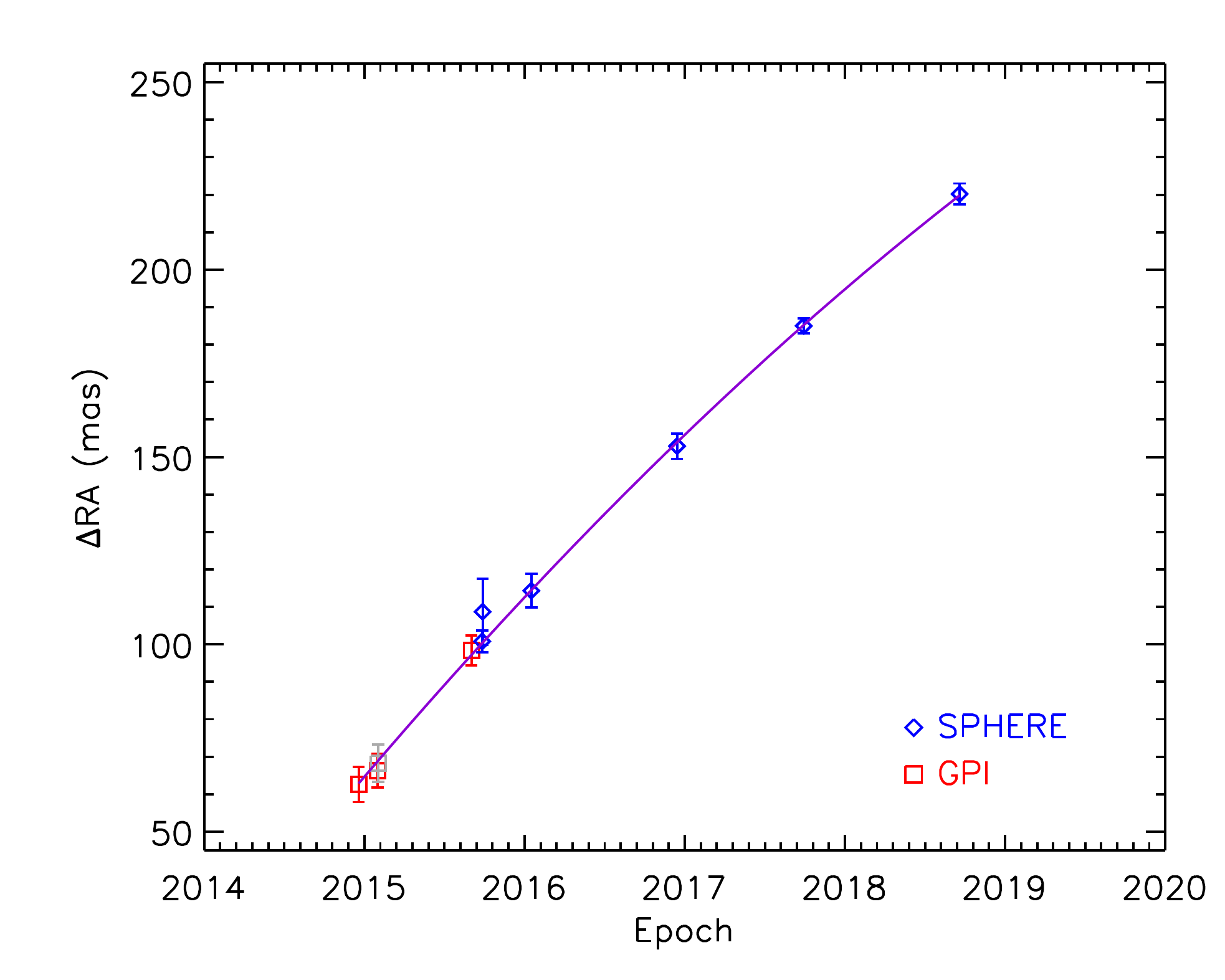}
\includegraphics[width=.42\textwidth]{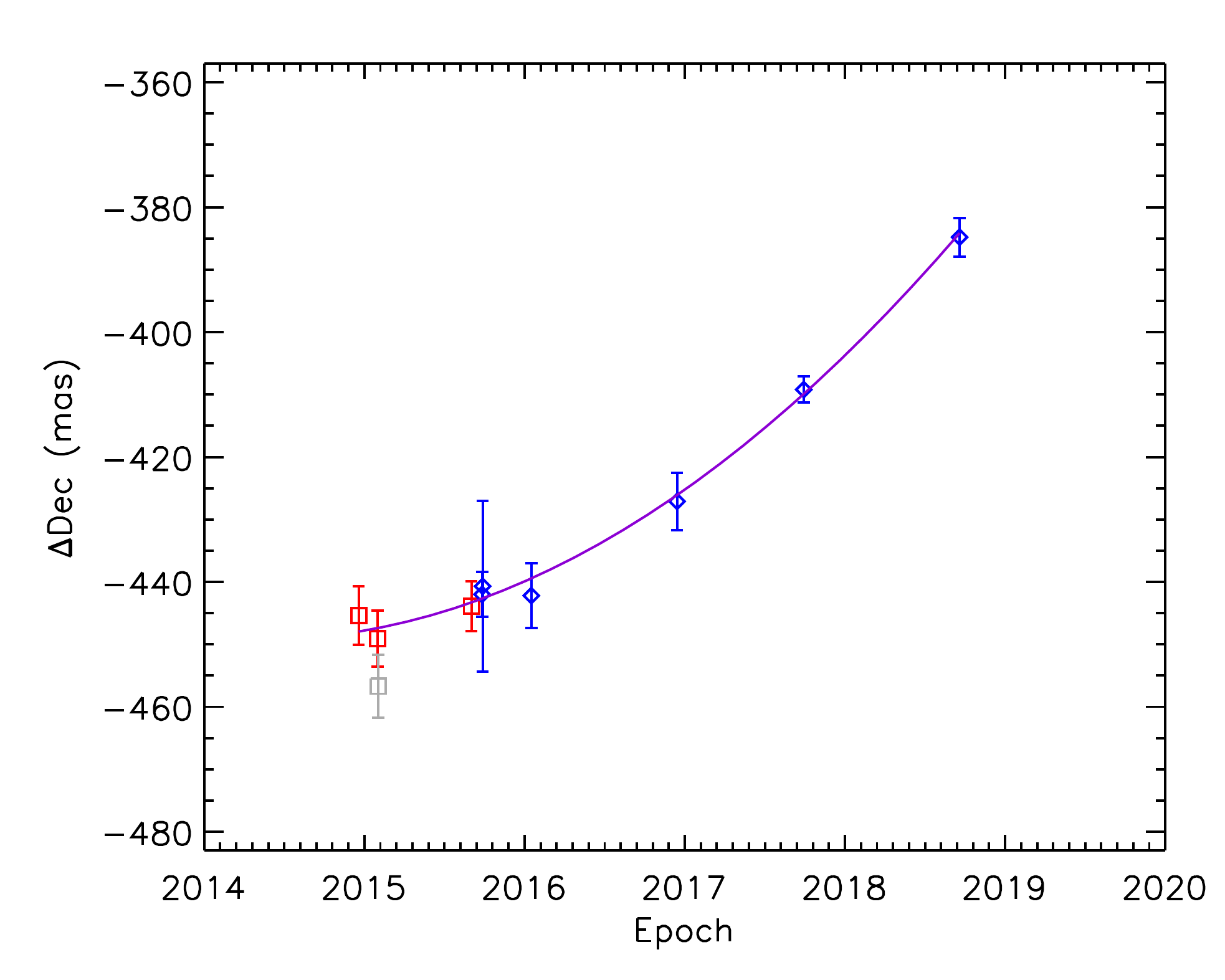}
\includegraphics[width=.42\textwidth]{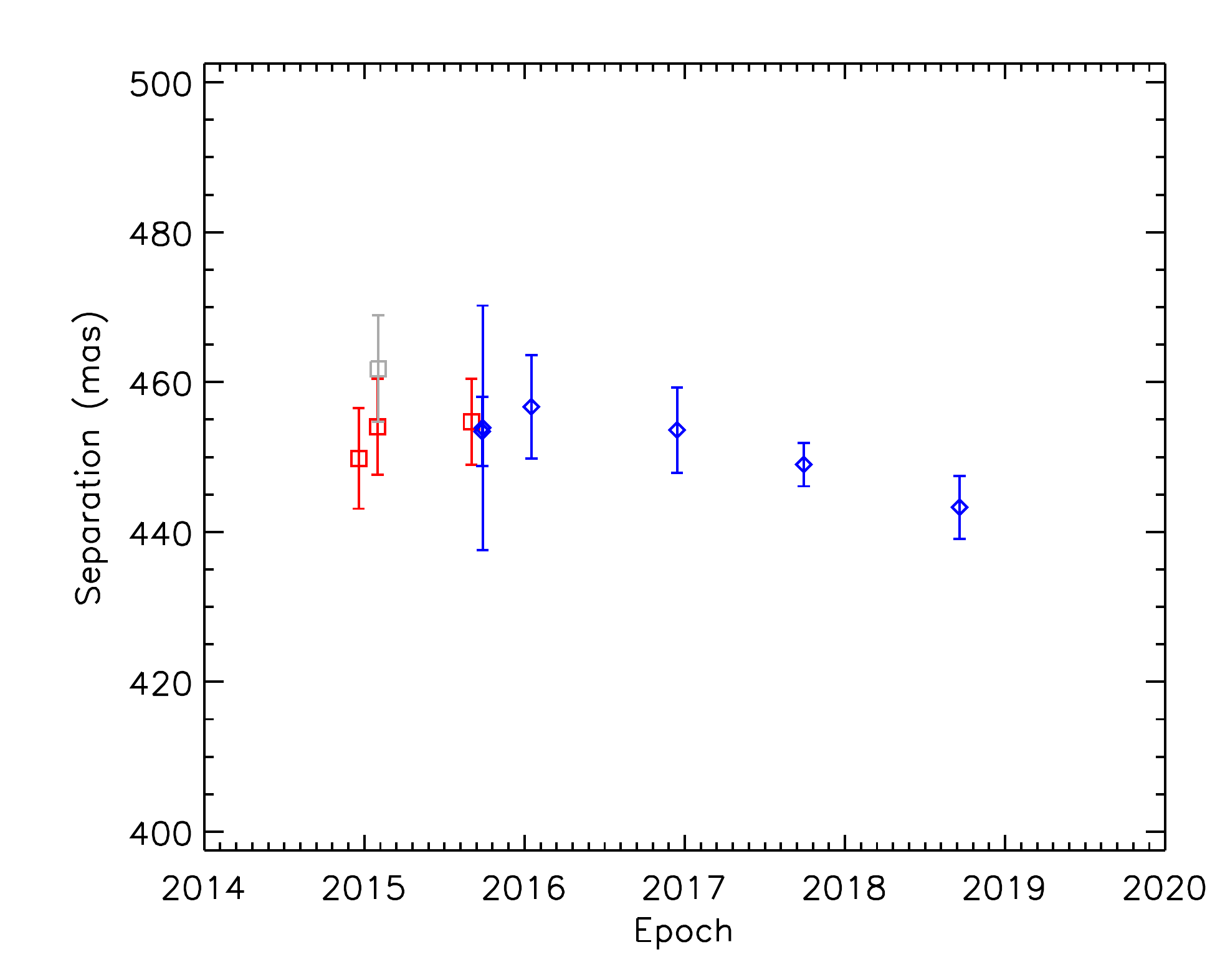}
\includegraphics[width=.42\textwidth]{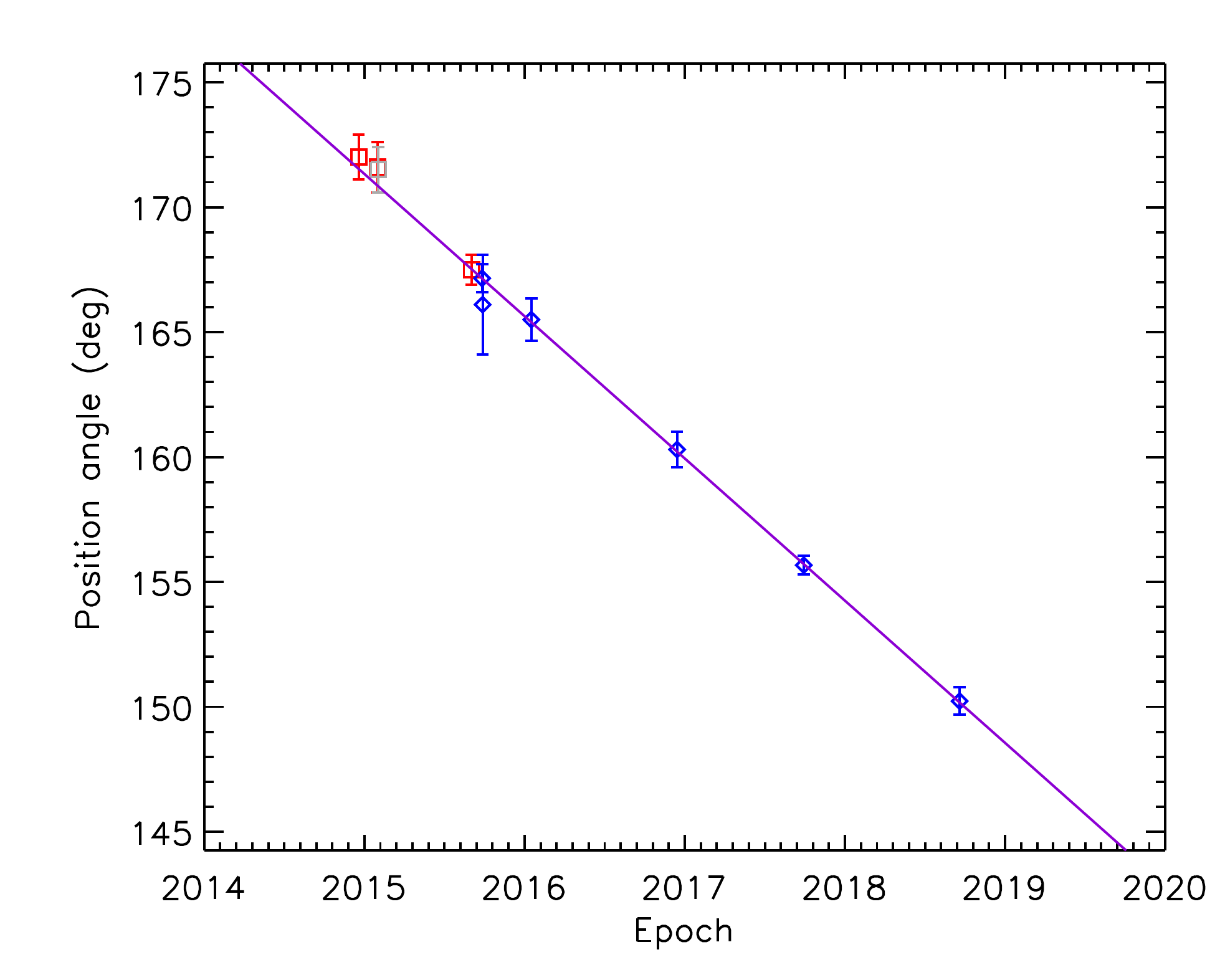}
\caption{{Temporal evolution of the relative right ascension (\textit{top left}), relative declination (\textit{top right}), separation (\textit{bottom left}) and declination (\textit{bottom right}) of 51~Eridani~b. With respect to Fig.~\ref{fig:51eri_astrometry}, the GPI astrometry is recalibrated by adding a position angle offset of 1.0$^{\circ}$ (see text). In the top panels, second-order polynomial fits on the SPHERE and GPI data points are also indicated to highlight the curvature in the planet motion (purple curves). The bottom-right panel shows a linear fit to the SPHERE data (purple line). The data point in light gray was not considered for the acceleration analysis and the orbital fits (see text).}}
\label{fig:51eri_seppa_time}
\end{figure*}

\section{Orbital motion}
\label{sec:astrometry} 

{The astrometry of the planet is given in Table~\ref{tab:astrometry}. The data} are represented in Fig.~\ref{fig:51eri_astrometry} with the GPI measurements reported by \citet{DeRosa2015}, who also revised the astrometry published in \citet{Macintosh2015}. The SPHERE data over three years confirm the orbital motion of the planet at a high significance: $\sim$119~mas at $\sim$30$\sigma$ in right ascension and $\sim$57~mas at $\sim$12$\sigma$ in declination. While there is a hint of a decrease in separation by $\sim$10~mas, the position angle clearly decreases at a rate of 5.7$\pm$0.2$^{\circ}$/yr. The trend in position angle is similar to the trend seen in the GPI data \citep{DeRosa2015}. The position angle variation is not compatible with the expectations for a {face-on circular orbit} ($\sim$10$^{\circ}$/yr, assuming a semi-major axis for the planet of 13~au and a stellar mass of 1.75~$M_\odot$), suggesting an inclined and/or eccentric orbit. The data also show signs of curvature, hinting at orbital inflexion (see below). 

The GPI data obtained in December 2014 and January 2015 show a discrepant increasing trend in separation (in particular, the separations measured on two consecutive nights in January 2015 are not included within the measurement uncertainties of each other {hence disagree at the 1-$\sigma$ level: 454.0\,$\pm$\,6.4~mas on January 30 and 461.8\,$\pm$\,7.1~mas on January 31}). \citet{Macintosh2015} noted that {the conditions for this last observation were average}. We also note a small systematic offset {in position angle} between the SPHERE and GPI data using two measurements obtained close in time in September 2015 {(the GPI point has PA\,=\,166.5\,$\pm$\,0.6$^{\circ}$, which is smaller by 1.1~$\sigma$ than the position angle of the SPHERE point which was obtained more than three weeks later; we should also expect a further displacement of about -0.38\,$\pm$\,0.02$^{\circ}$ in position angle due to the orbital motion of the planet over this elapsed time)}. This offset is likely due to systematic uncertainties related to differences in the astrometric calibration of the instruments (see Appendix~\ref{sec:gpispherecomparison}). It is well accounted for by an offset in the measured position angles of 1.0$\pm$0.2$^{\circ}$. The inclusion of the recalibration uncertainty in the GPI measurement uncertainties has negligible effects.

We show the recalibrated GPI data and the SPHERE measurements in Fig.~\ref{fig:51eri_seppa_time}, which represents the temporal evolution of the {relative right ascension, relative declination, separation, and position angle}. The curvature of the SPHERE data suggested from Fig.~\ref{fig:51eri_astrometry} is better seen in the top panels of Fig.~\ref{fig:51eri_seppa_time}. We also show in these panels second-order polynomial fits to {all the SPHERE and GPI data except for the GPI data point taken on 2015 January 31 due to its discrepant separation with respect to the other GPI data points. {{{The separation measured for this epoch is 461.8~mas whereas the other data points have separations smaller than 455~mas (upper limit of 460.4~mas at 1$\sigma$). Even when increasing the error bars on this data point to include the other GPI measurements, the LSMC orbital fit is affected with respect to a fit where this data point is excluded and shows a stronger paucity in low-eccentricity orbits}}}.} {Second-order polynomial fits provide significantly better unreduced chi-square goodness-of-fit parameters (1.4 for $\Delta$RA vs. time and 1.0 for $\Delta$Dec vs. time) with respect to linear fits (7.4 and 12.2, respectively).} We followed the approach of \citet{Konopacky2016a} to test if acceleration is detected (the measured acceleration plus its uncertainty at 3$\sigma$ shall stay negative). From the above-mentioned second-order polynomial fits, we estimated the cartesian components of the acceleration and converted them into radial and tangential components. The radial acceleration component is {$-$4.03$\pm$1.34~mas.yr$^{-2}$, which implies that acceleration is detected at the 3.0$\sigma$ level.} New measurements should help to confirm the acceleration estimate, and measure it with better accuracy.

\begin{figure*}[t]
\centering
\includegraphics[width=.96\textwidth]{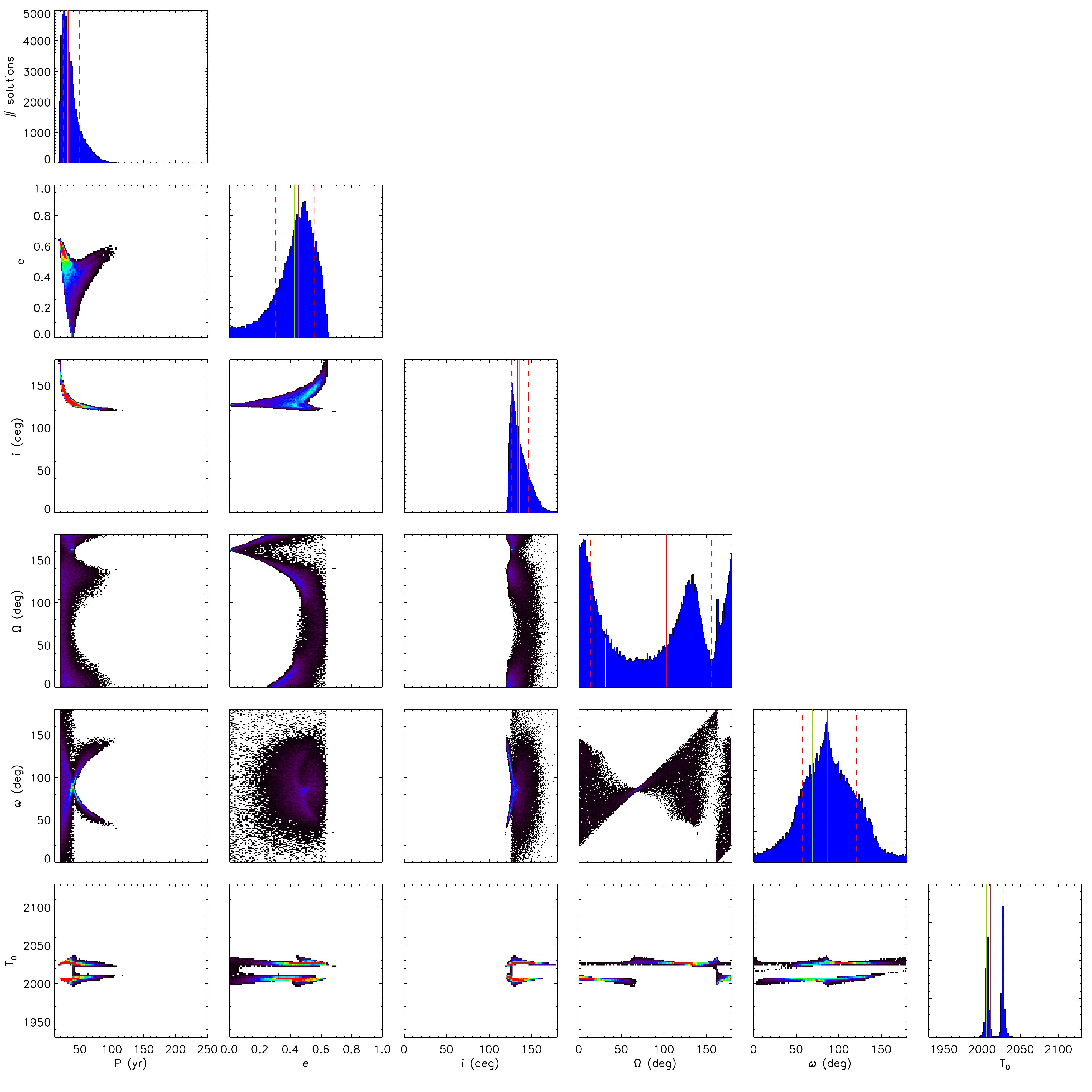}
\caption{{LSMC distributions of the six Campbell orbital elements for all the fitted solutions with $\chi_{\rm{red}}^2$\,<\,2 among 1\,000\,000 random trials. The diagrams displayed on the diagonal from top left to lower right represent the 1D histogram distributions for the individual elements. The off-diagonal diagrams show the correlations between pairs of orbital elements. The linear color-scale in the correlation plots accounts for the relative local density of orbital solutions. In the histograms, the green solid line indicates the best $\chi^2$ fitted solution, the red solid line shows the 50\% percentile value, and the red dashed lines the interval at 68\%.}}
\label{fig:51eri_cornerplot}
\end{figure*}

\section{Orbital analysis}
\label{sec:orbit}

\begin{figure*}[t]
\centering
\includegraphics[height=.28\textwidth]{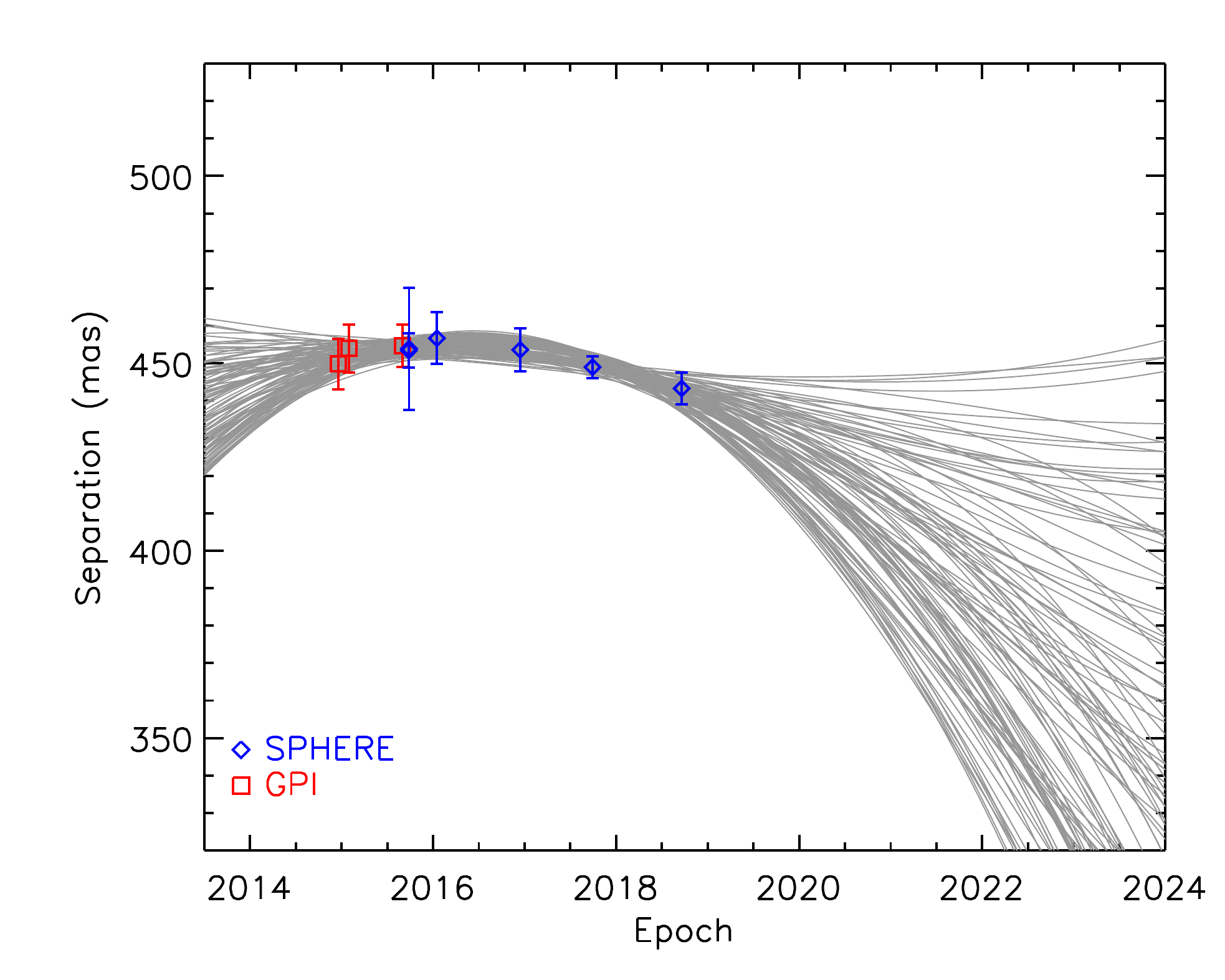}
\includegraphics[height=.28\textwidth]{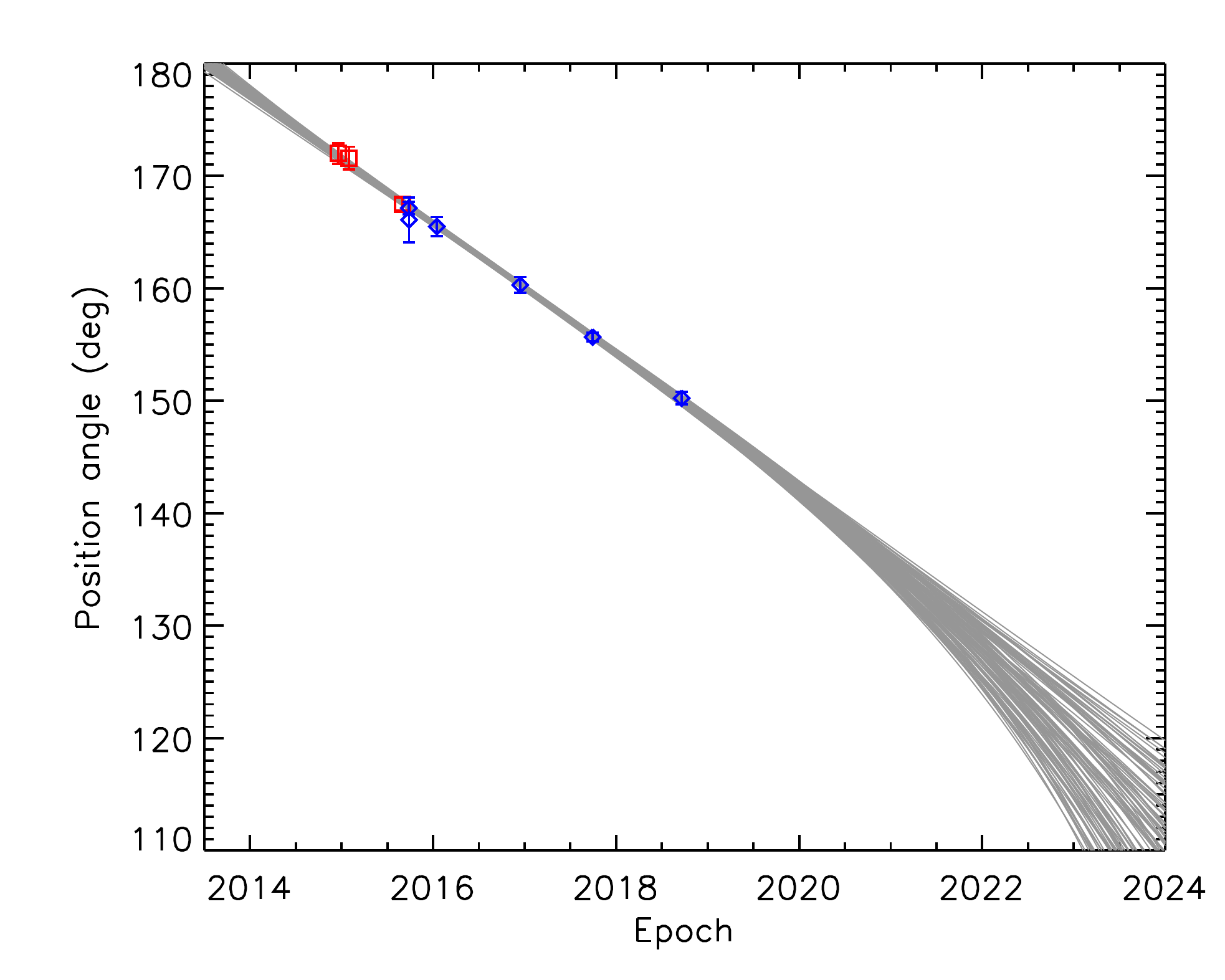}
\includegraphics[height=.28\textwidth]{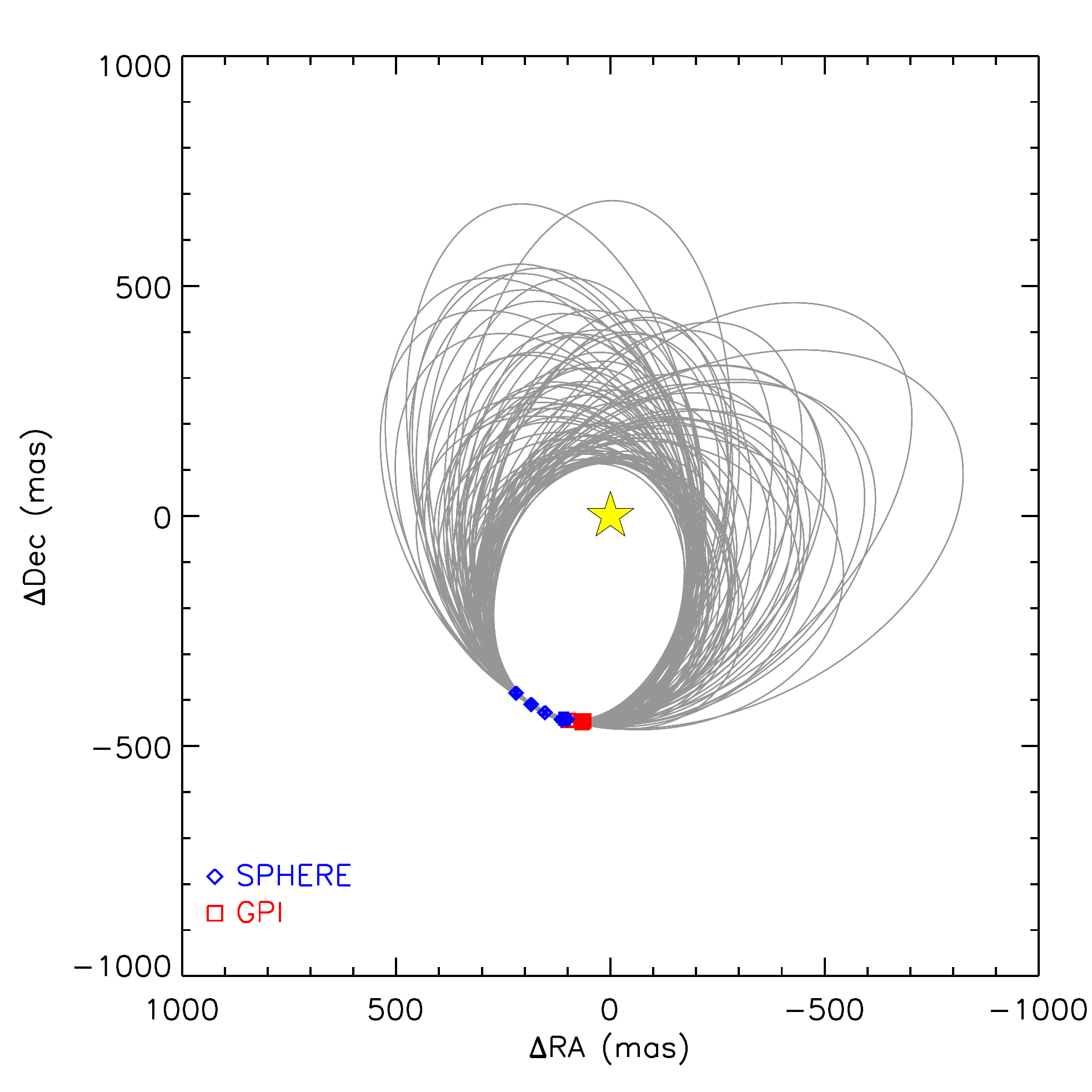}
\caption{{Predictions for the separation (\textit{left}) and position angle (\textit{middle}) of 51 Eridani b for 100 randomly selected orbital solutions in Fig.~\ref{fig:51eri_cornerplot}. The {right} panel displays the orbits in the plane of the sky.}}
\label{fig:51eri_seppa_time_predictions}
\end{figure*}

\subsection{Determination of the orbital parameters}

We assumed for the system the distance estimated from the \textit{Gaia} parallax and a total mass of 1.75~$M_{\sun}$ \citep{Simon2011}. We also make the assumption that the GJ~3305AB binary does not dynamically disturb the orbit of the planet. \citet{Montet2015} showed that given the wide separation of the binary and the young age of the system, it appears unlikely that Kozai-Lidov oscillations {\citep{Kozai1962, Lidov1962}} could have had the time to {disturb} the semi-major axis of the planet (typical timescale of 200~Myr for a perturber on a circular orbit). Nevertheless, moderate changes in the inclination and eccentricity cannot be excluded \citep{Fabrycky2007}. Quicker dynamical effects (secular precession {due to, e.g.,} an unseen inner companion) could suppress Kozai-Lidov oscillations. On the other hand, assuming the criterion in \citet{Holman1999}, we could also expect that the current orbit remains unchanged despite the wide binary.

We first used a least-squares Monte Carlo (LSMC) procedure to fit the SPHERE {and GPI astrometry} \citep[][Appendix~\ref{sec:orbfamiliesppties}]{Esposito2013,Maire2015}. We also performed complementary analyses using a Markov-chain Monte Carlo (MCMC) procedure \citep[][Appendix~\ref{sec:mcmcorbit}]{Chauvin2012} and the Bayesian rejection sampling approach OFTI \citep[][Appendix~\ref{sec:oftiorbit}]{Blunt2017}.

We used the LSMC and OFTI methods to {analyze} the effects of the inclusion of the GPI astrometric points on the parameter distributions. Adding the GPI data strengthens short-period and eccentric orbits. We also used these two approaches to {test the effect of a different initial eccentricity distribution on the resulting eccentricity distribution,} given that the analyses favor eccentric orbits over circular orbits. We used a distribution that gives more weight to low-eccentricity orbits, similar to the fit to the eccentricity distribution of radial-velocity planets in \citet{Nielsen2008}. The {resulting eccentricity distributions} have similar shapes using both types of initial distributions. {Finally, we checked that the uncertainty on the stellar mass ($<$3\%) has negligible effects on the parameter distributions.}

Further constraints could also be obtained with detection limits from archival high-contrast imaging data by rejecting orbits predicting overly large separations for the planet in the past. Unfortunately, the current detection limits are not deep enough to provide useful constraints \citep{Heinze2010, Biller2013, Rameau2013a, Hagan2018, Stone2018}.

\subsection{Parameter intervals and correlations}
\label{sec:paramcorr}

The LSMC distributions and intervals of the orbital parameters are shown in Fig.~\ref{fig:51eri_cornerplot} and Table~\ref{tab:orbparams}. The shapes of the distributions and the parameter intervals are {broadly} similar to those obtained with the MCMC and OFTI approaches. The three $T_0$ distributions show two peaks around $\sim$2005 and $\sim$2025. {The three $\Omega$ distributions display two broad peaks around $\sim$10$^{\circ}$ and $\sim$130$^{\circ}$. Nevertheless, we note some differences in the detailed shape of the distributions. The LSMC eccentricity distribution does not show a high-eccentricity tail (correlated with a long-period tail) as seen in the MCMC and OFTI distributions. We verified that this is due to the correction for the LSMC fitted orbits of the bias on the time at periastron passage for eccentric orbits \citep{Konopacky2016a}. The LSMC distribution for $i$ has a more pronounced peak toward values smaller than $\sim$135$^{\circ}$ with respect to the MCMC and OFTI distributions. This is caused by a larger number of low-eccentricity orbits with $\Omega$ around $\sim$160$^\circ$ in the LSMC fitted orbits. The LSMC distribution for $\Omega$ exhibits a deeper dip between the two broad peaks mentioned above with respect to the MCMC and OFTI distributions. This feature is related to a paucity of orbits with $\Omega$ around $\sim$65$^{\circ}$ and low to moderate eccentricities (up to $\sim$0.4) in the LSMC fitted orbits. The MCMC distribution for $\omega$ appears sharper than the LSMC and OFTI distributions.}

\begin{table}[t]
\caption{{Preliminary orbital parameters of 51~Eridani~b}}
\label{tab:orbparams}
\begin{center}
\begin{tabular}{l c c c c c}
\hline\hline
Parameter & Unit & Median & Lower & Upper & $\chi^2_{\rm{min}}$ \\
\hline
$P$ & yr & 32 & 23 & 49 & 30 \\
$a$ & au & 12 & 10 & 16 & 12 \\
$e$ & & 0.45 & 0.30 & 0.55 & 0.43 \\
$i$ & $^{\circ}$ & 133 & 126 & 147 & 135 \\
$\Omega$ & $^{\circ}$ & 103 & 13 & 156 & 17 \\
$\omega$ & $^{\circ}$ & 87 & 57 & 121 & 69 \\
$T_0$ & & 2011 & 2006 & 2027 & 2006 \\
\hline
\end{tabular}
\end{center}
\tablefoot{The parameters are the period, semi-major axis, eccentricity, inclination, longitude of node (mod 180$^{\circ}$), argument of periastron passage (mod 180$^{\circ}$), and time at periastron passage. {The median value is the 50\% value, the lower and upper values are the lower and upper bounds of the 68\% interval, and the $\chi^2_{\rm{min}}$ value is the best-fit value.}}
\end{table}

When comparing our results obtained with the OFTI approach with those of \citet{DeRosa2015}, who employed a similar method, we note that most of the parameters are {better defined in our analysis}{. {The 68\% interval for the longitude of node in \citet{DeRosa2015} was derived by wrapping the values to the range 30--120$^{\circ}$, whereas we consider the full [0;180$^{\circ}$] range. These authors also wrapped their distribution of the time of periastron passage to the range 1995--1995+$P$, whereas our distributions extend to previous epochs. Nevertheless, we verified that applying a similar wrapping has a negligible effect on the derived $T_0$ ranges in our analysis.}} We derive a 68\% interval for the inclination of {126--147$^{\circ}$}, which is slightly better constrained with respect to the range of 125--153$^{\circ}$ in \citet{DeRosa2015}. Thus, we confirm after \citet{DeRosa2015} that the orbital plane of the planet {cannot be} coplanar with the orbital plane of the wide-separated binary GJ~3305AB \citep[$i$\,=\,92.1$\pm$0.2$^{\circ}$,][]{Montet2015}. Our eccentricity interval is {0.33--0.57} at 68\%, which suggests eccentric orbits. This feature is related to our more extended dataset with respect to \citet{DeRosa2015} (Appendix~\ref{sec:oftiorbit}). Nevertheless, orbits with low to moderate eccentricities are not formally excluded. {Low-eccentricity orbits are characterized by parameters around $\Omega$\,$\sim$\,160$^{\circ}$, $i$\,$\sim$\,130$^{\circ}$, and $P$\,$\sim$\,35--40~yr.} New data covering a larger fraction of the orbit of the planet are needed to verify any suggestion of large eccentricities.

Figure~\ref{fig:51eri_seppa_time_predictions} shows a random sample of fitted orbits from the LSMC analysis. {In particular, low-eccentricity orbits predict steeper decreases of the separation and position angle in the coming years.} {If new data taken in the next couple of years show a weak decrease of the separation and still follow the linear trend in position angle seen with the current data, this will rule out low-eccentricity orbits.}

\subsection{An unseen inner companion that could bias the  eccentricity of the planet?}
\label{sec:pearceanalysis}
\begin{figure}[t]
\centering
\includegraphics[trim=14mm 0mm 14mm 10mmm,clip,width=0.4\textwidth]{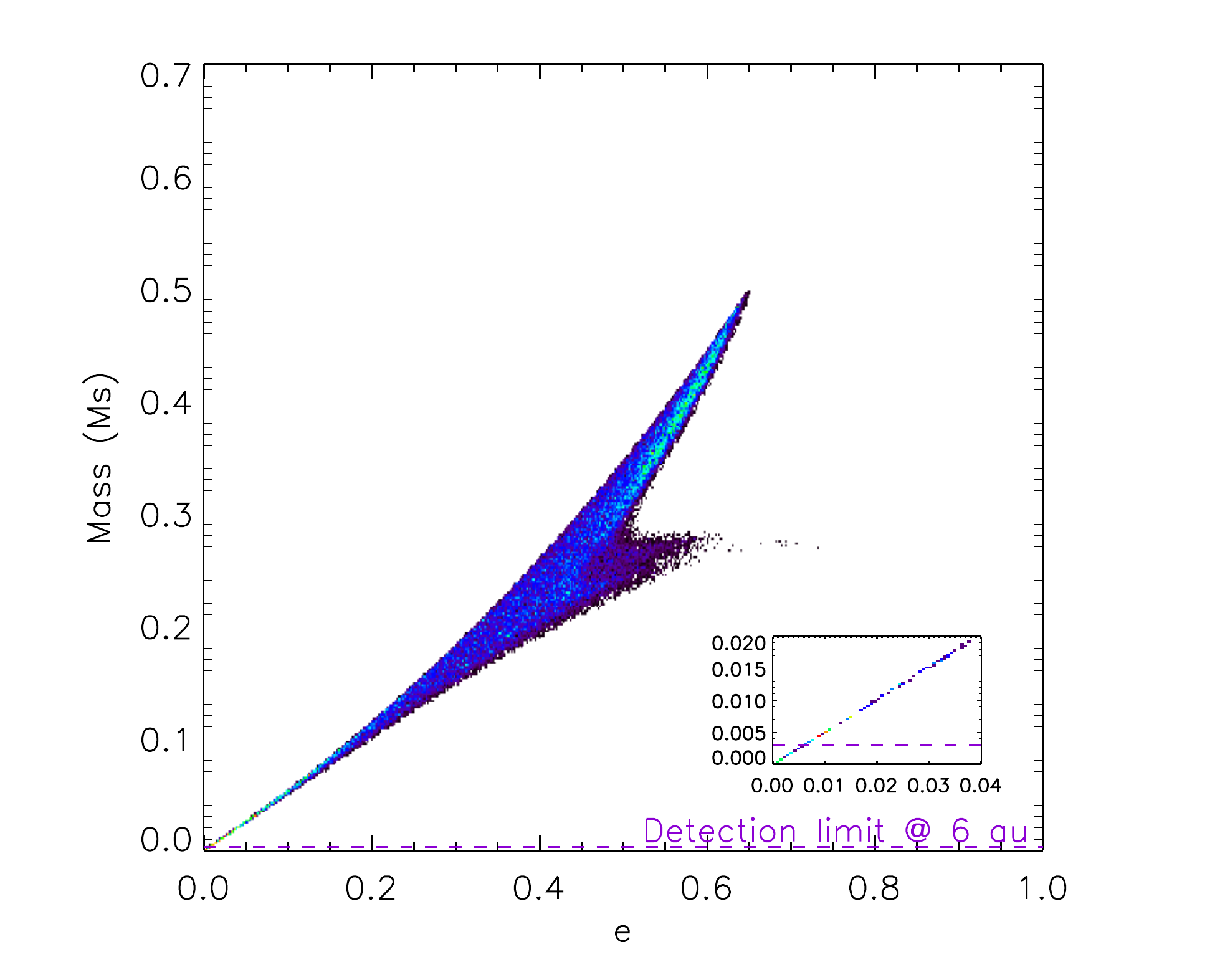}
\caption{{Minimum mass (in solar masses) of an unseen inner companion on a circular orbit that could bias the eccentricity measured for 51~Eridani~b compared to the SPHERE/IRDIS detection limit at 6~au (see text). The inset provides a zoom at low eccentricities and masses to better show the detection limit.}}
\label{fig:pearcetest}
\end{figure}

Finally, we used the methods in \citet{Pearce2014} to test the scenario of an unseen inner low-mass companion which could bias the eccentricity of 51~Eridani~b toward large values due to the orbital motion that the unseen companion induces on the host star around the center of mass of the system. We considered the case where this putative inner companion lies on a circular orbit. For this, we used the period and eccentricity distributions in Fig.~\ref{fig:51eri_cornerplot}. Figure~\ref{fig:pearcetest} shows the {minimum mass} of a putative inner companion as a function of the eccentricity of the planet. Such a companion would lie at an angular separation of $\sim$0.21$''${ ($\sim$6~au)}. By comparing these masses to the SPHERE/IRDIS mass limit of $\sim$3~$M_{\rm{J}}$ measured at this separation \citep{Samland2017} according to the atmospheric and evolutionary models of \citet{Baraffe2015, Baraffe2003}, we can conclude that if 51~Eridani~b has a nonzero eccentricity, this eccentricity is genuine and does not result from an unseen low-mass inner companion. {The companion could in theory lie at a larger separation but its mass would be larger, hence its detection would be even easier.}

\section{Conclusions}

We presented VLT/SPHERE observations over three years of the young giant exoplanet 51~Eridani~b to further characterize its orbital motion and parameters. The planet moved by $\sim$130~mas over this elapsed time with hints of orbital curvature and a decreasing trend in its separation to the star. We compared the results of three orbital fitting approaches based on LSMC, MCMC, and Bayesian rejection sampling and found {similar distribution shapes for all parameters}. With respect to the study of \citet{DeRosa2015}, our orbital analysis based on a similar Bayesian rejection sampling approach provides {narrower} ranges {for the orbital parameters. The time at periastron passage and the longitude of node exhibit bimodal distributions, the ambiguity being related to the non-detection of changes in the orbital speed of the planet}. We derived an inclination range of {126--147$^{\circ}$}, which is slightly {narrower} than the 125--153$^{\circ}$ range derived by \citet{DeRosa2015}. We note that {the orbital inclination of the planet is compatible with an orbit lying in the stellar equatorial plane (the stellar rotation axis has an inclination of 134--144$^{\circ}$; see Appendix~\ref{sec:photometry}) or offset by less than {18$^{\circ}$}}. Given that the star is expected to host a debris disk, this might suggest a coplanar planet--disk configuration and dynamical interactions. {Further astrometric monitoring will help to solve for the ambiguities in the time at periastron passage and the longitude of node as well as refine the orbital inclination of the planet and the analysis of the system spin-orbit.}

{Our orbital analysis suggests an eccentric orbit for the planet with a 68\% interval of {0.30--0.55}. If the eccentricity of 51~Eridani~b is indeed genuine, this may hint at dynamical interactions between the planet and another body in the system to produce such a large eccentricity. This putative additional body could be an unseen inner or outer planet, although we note that the current imaging detection limits are relatively deep \citep[$>$4~$M_{\rm{J}}$ beyond 5~au and $>$2~$M_{\rm{J}}$ beyond 9~au assuming hot-start models;][]{Samland2017}. Another possibility would be gravitational perturbations from GJ~3305AB, such as Kozai-Lidov oscillations, but this scenario may face timescale issues because of the large separation of the binary and the system youth. \citet{Fabrycky2007} predict that a close-in giant planet experiencing Kozai-Lidov oscillations from a distant binary companion to its host star will typically have an orbit that is misaligned with the stellar equatorial plane. Our analysis favors alignment or misalignment within 18$^{\circ}$ for the spin-orbit of the 51~Eridani star--planet system. Interactions of the planet with the circumstellar disk could also be a possible mechanism. Although this kind of interaction is usually thought to dampen the eccentricity of a planet, simulations have shown that for massive giant planets (>4--5~$M_{\rm{J}}$) lying near the disk, midplane (<10$^{\circ}$) interactions with a protoplanetary disk increase their eccentricity \citep{Papaloizou2001, Kley2006, Bitsch2013}. The current mass estimate of 51~Eridani~b is $\sim$2--4~$M_{\rm{J}}$ assuming hot-start models, but it could be as large as 12~$M_{\rm{J}}$ assuming warm-start models \citep{Samland2017}.}

Further astrometric monitoring in the next 3--4 years will be critical to confirm at a higher significance the curvature in the motion of the planet, determine if it is accelerating or decelerating on its orbit, and further constrain its orbital parameters. It will also be critical for preparing future observations. If the angular separation of  51~Eridani~b strongly decreases, this might prevent follow-up investigations near its periastron passage with SPHERE, GPI, and {the \textit{James Webb Space Telescope}} to better constrain its orbital and atmospheric properties and {leave} such observations feasible with {Extremely Large Telescope} instruments only.

In addition to further orbital follow-up studies, resolved images of the host star debris disk will be valuable to determine if the planet orbits in the disk plane by providing the disk orientation and, if the data confirm the two disk belts inferred from the stellar SED, if it dynamically shapes these belts by providing their radial extent. Such information would also help to better characterize the orbital period and eccentricity of 51~Eridani~b.

\begin{acknowledgements}
     {We thank an anonymous referee for a constructive report that helped to improve the manuscript}. The authors thank the ESO Paranal Staff for support in conducting the observations and Philippe Delorme, Eric Lagadec, and Nad\`ege Meunier (SPHERE Data Center) for their help with the data reduction. {We also thank Remko Stuik for discussions about the analysis of the MASCARA data and Sarah Blunt, Henry Ngo, Jason Wang, Isabel Angelo, Devin Cody, Robert De Rosa, and Logan Pearce for making the \texttt{orbitize} code publicly available}. We acknowledge financial support from the Programme National de Planétologie (PNP) and the Programme National de Physique Stellaire (PNPS) of CNRS-INSU. This work has also been supported by a grant from the French Labex OSUG@2020 (Investissements d'avenir -- ANR10 LABX56). The project is supported by CNRS, by the Agence Nationale de la Recherche (ANR-14-CE33-0018). D.M. was supported by the South Carolina Space Grant Consortium. J.C. was supported by the Fulbright Colombia program as well as the South Carolina Space Grant Consortium. This work has made use of the SPHERE Data Centre, jointly operated by OSUG/IPAG (Grenoble), PYTHEAS/LAM/CeSAM (Marseille), OCA/Lagrange (Nice) and Observatoire de Paris/LESIA (Paris). This research made use of the SIMBAD database and the VizieR Catalogue access tool, both operated at the CDS, Strasbourg, France. The original description of the VizieR service was published in Ochsenbein et al. (2000, A\&AS 143, 23). This research has made use of NASA's Astrophysics Data System Bibliographic Services. SPHERE is an instrument designed and built by a consortium consisting of IPAG (Grenoble, France), MPIA (Heidelberg, Germany), LAM (Marseille, France), LESIA (Paris, France), Laboratoire Lagrange (Nice, France), INAF -- Osservatorio di Padova (Italy), Observatoire astronomique de l'Universit\'e de Gen\`eve (Switzerland), ETH Zurich (Switzerland), NOVA (Netherlands), ONERA (France), and ASTRON (Netherlands), in collaboration with ESO. SPHERE was funded by ESO, with additional contributions from the CNRS (France), MPIA (Germany), INAF (Italy), FINES (Switzerland), and NOVA (Netherlands). SPHERE also received funding from the European Commission Sixth and Seventh Framework Programs as part of the Optical Infrared Coordination Network for Astronomy (OPTICON) under grant number RII3-Ct-2004-001566 for FP6 (2004--2008), grant number 226604 for FP7 (2009--2012), and grant number 312430 for FP7 (2013--2016).

\end{acknowledgements}

%
   \bibliographystyle{aa} 
   \bibliography{biblio} 
%

\begin{appendix}

\section{Comparison of astrometric measurements from different algorithms}
\label{sec:otheralgos}

\begin{table}[h]
\caption{{SPHERE astrometry relative to the star of 51~Eridani~b obtained with SDI+TLOCI and SDI+ANDROMEDA.}}
\label{tab:astrometryotheralgos}
\begin{center}
\begin{tabular}{l c c c}
\hline\hline
Epoch & Spectral band & $\rho$ (mas) & PA ($^{\circ}$) \\
\hline
\multicolumn{4}{c}{SDI+TLOCI} \\
\hline
2015.74 & $K1$ & 453.4$\pm$4.4 & 167.15$\pm$0.55 \\ 
2015.74 & $H$ & 453.9$\pm$15.9 & 166.1$\pm$2.0 \\
{2016.04} & $H2$ & 456.7$\pm$6.6 & 165.50$\pm$0.83 \\
2016.95 & $H2$ & 453.6$\pm$5.7 & 160.30$\pm$0.72 \\
2017.74 & $K1$ & 449.0$\pm$2.9 & 155.67$\pm$0.37 \\
2018.72 & $K1$ & 443.3$\pm$4.2 & 150.23$\pm$0.54 \\
\hline
\multicolumn{4}{c}{SDI+ANDROMEDA} \\
\hline
{2015.74} & $K1$ & 448.6$\pm$1.4 & 167.45$\pm$0.06 \\
2015.74 & $H$ & 467.4$\pm$2.9 & 167.09$\pm$0.07 \\
2016.04\tablefootmark{a} & $H2$ & -- & -- \\
{2016.95} & $H2$ & 456.1$\pm$1.6 & 160.06$\pm$0.06 \\
{2017.74} & $K1$ & 447.9$\pm$1.3 & 155.80$\pm$0.04 \\
{2018.72} & $K1$ & 439.0$\pm$1.2 & 150.09$\pm$0.03 \\
\hline
\end{tabular}
\end{center}
\tablefoot{The uncertainties are from the measurement procedure only and are given at 1$\sigma$. \\
\tablefootmark{a}{It was not possible to extract the astrometry of the planet.}
}
\end{table}

\section{Comparison of the SPHERE and GPI astrometry without GPI/SPHERE recalibration}
\label{sec:gpispherecomparison}

\begin{figure*}[t]
\centering
\includegraphics[width=.4\textwidth]{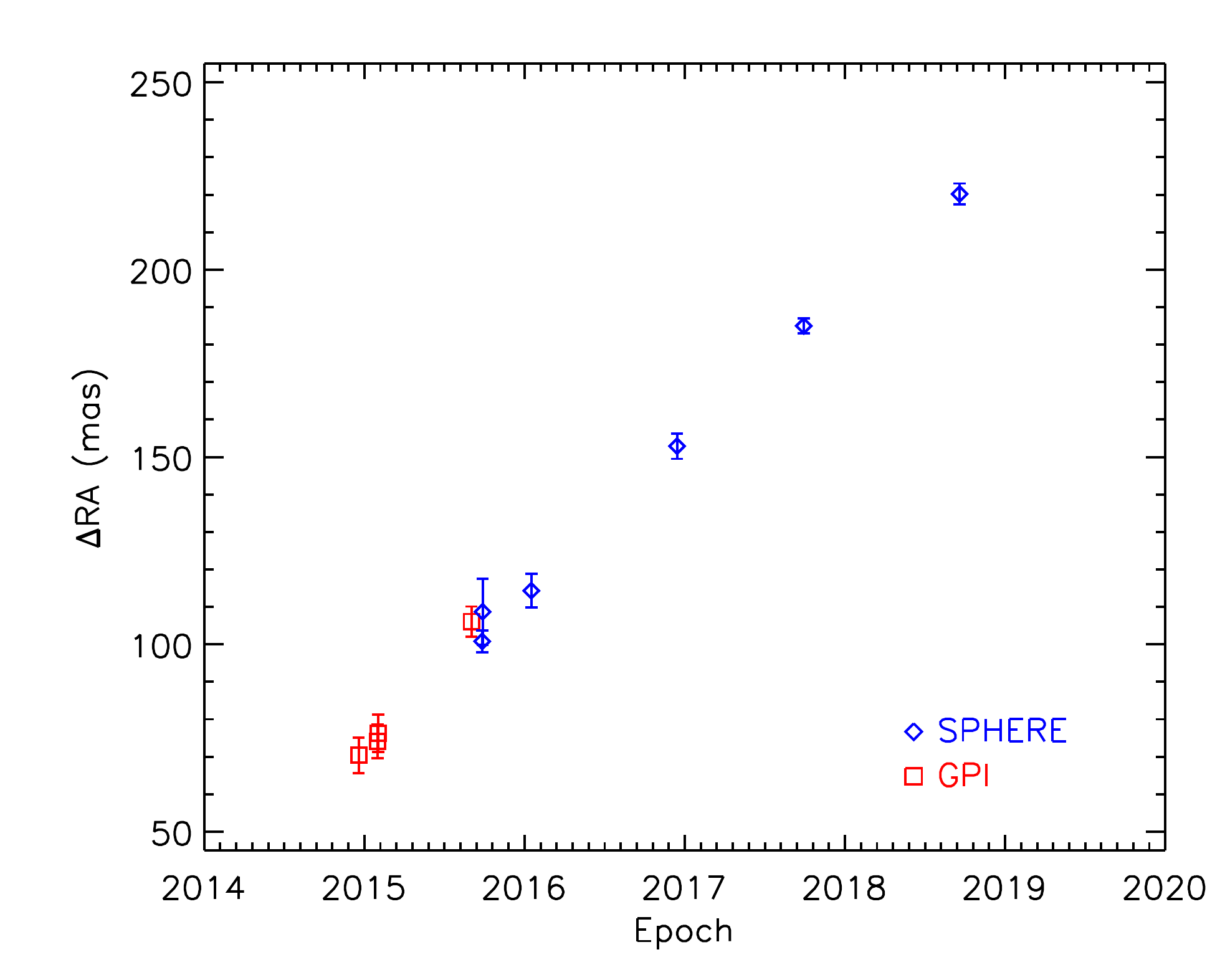}
\includegraphics[width=.4\textwidth]{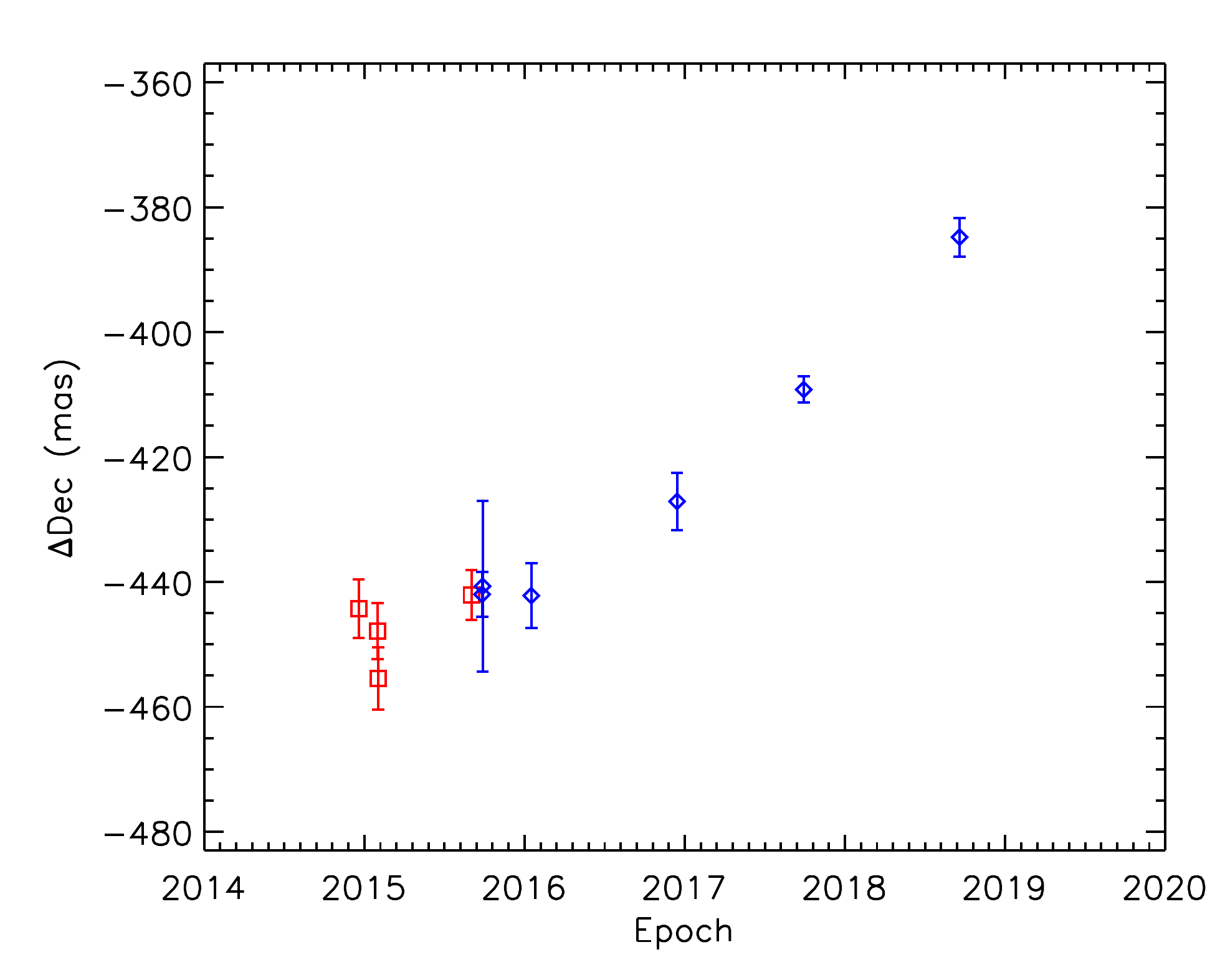}
\includegraphics[width=.4\textwidth]{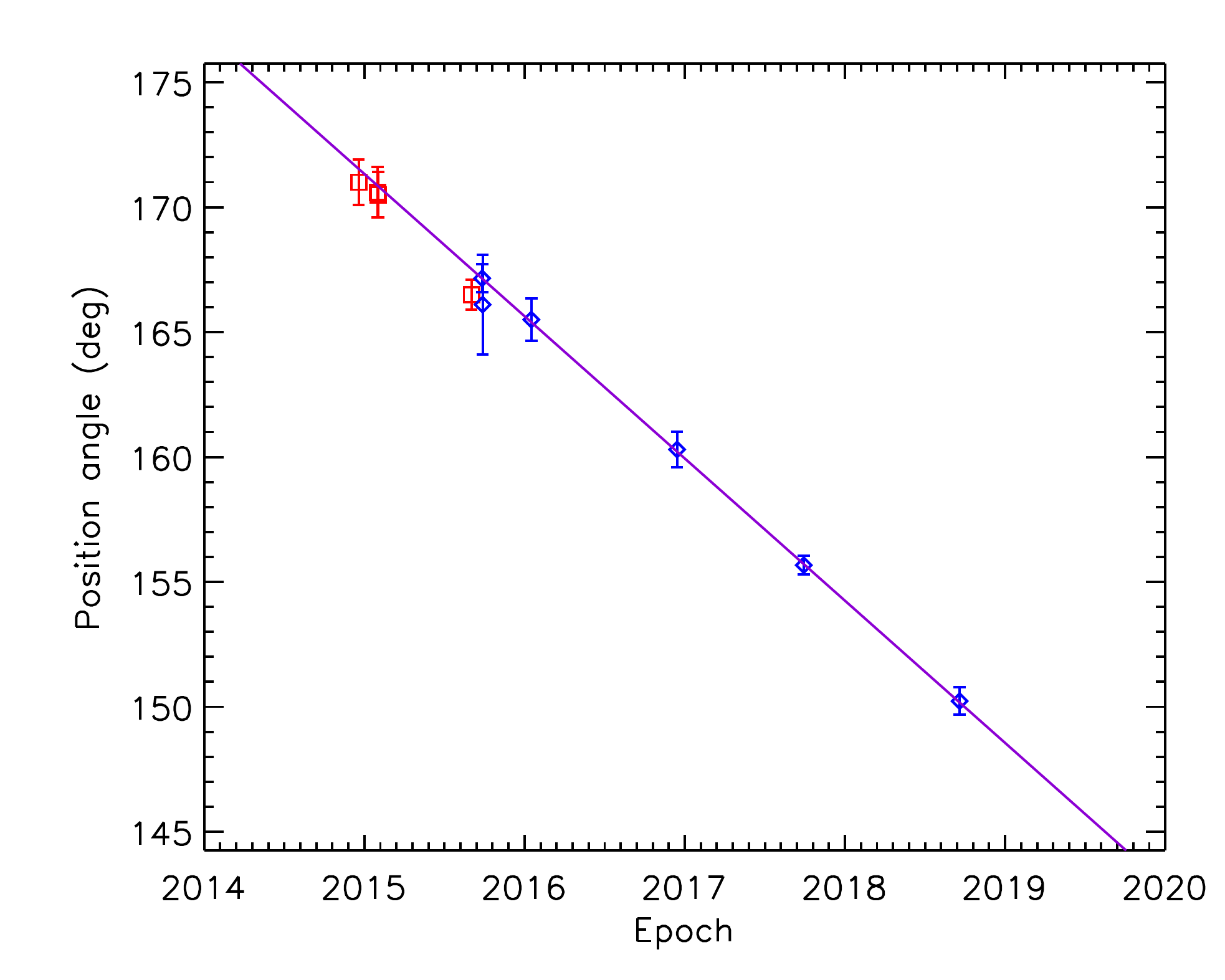}
\caption{{Temporal evolution of the relative right ascension (\textit{top left}), relative declination (\textit{top right}), and position angle (\textit{bottom}) of 51~Eridani~b. The GPI astrometry is taken from \citet{DeRosa2015} without applying a recalibration of the position angle measurements (see Appendix~\ref{sec:gpispherecomparison}). The bottom-right panel shows a linear fit to the SPHERE data (purple line).}}
\label{fig:51eri_seppa_time_norecal}
\end{figure*}

{Figure~\ref{fig:51eri_seppa_time_norecal} shows further comparisons of all the SPHERE astrometry and the GPI astrometry reported in \citet{DeRosa2015} without applying any recalibration of the latter on the SPHERE data. While we do not see any clear GPI/SPHERE offset in the separations (Fig.~\ref{fig:51eri_seppa_time}), we note that the position angle of the GPI point taken on 2015 September 1 is  0.65$^{\circ}$ (1.1 times the measurement uncertainty) smaller than the position angle of a SPHERE point taken more than three weeks later, on 2015 September 25. However, we expect  an additional decrease in position angle of $\sim$0.38$\pm$0.02$^{\circ}$ between the GPI and SPHERE epochs due to the orbital motion of the planet, meaning that the actual offset between these two measurements is possibly $\sim$1$^{\circ}$ (1.7 times the measurement uncertainty).}

{We did not consider for the comparison the SPHERE point taken on 2015 September 26 because the position angle measured at this epoch has significantly larger uncertainties and deviates from a linear fit matching all the other SPHERE data points well (purple line in the bottom panel of Fig.~\ref{fig:51eri_seppa_time_norecal}). Contrary to the other SPHERE datasets, this dataset was not obtained with the dual-band imaging mode of IRDIS that allows for simultaneous imaging in two spectral bands in and out of a methane absorption band. Therefore, SDI could not be used in the image post-processing to attenuate fast quasi-static stellar speckles which are not attenuated with angular differential imaging, resulting in a poorer detection of the planet. Furthermore, we did not consider other GPI and SPHERE data points because they were taken {at large time intervals}. We note that the GPI position angle measurements exhibit a steeper slope with respect to the SPHERE data although the measured uncertainties are large ($-$6.7$\pm$1.3$^{\circ}$) and include the slope value derived from the SPHERE data ($-$5.7$\pm$0.2$^{\circ}$).}

\begin{figure}
\centering
\includegraphics[trim = 5mm 2mm 5mm 160mm, width=.44\textwidth,clip]{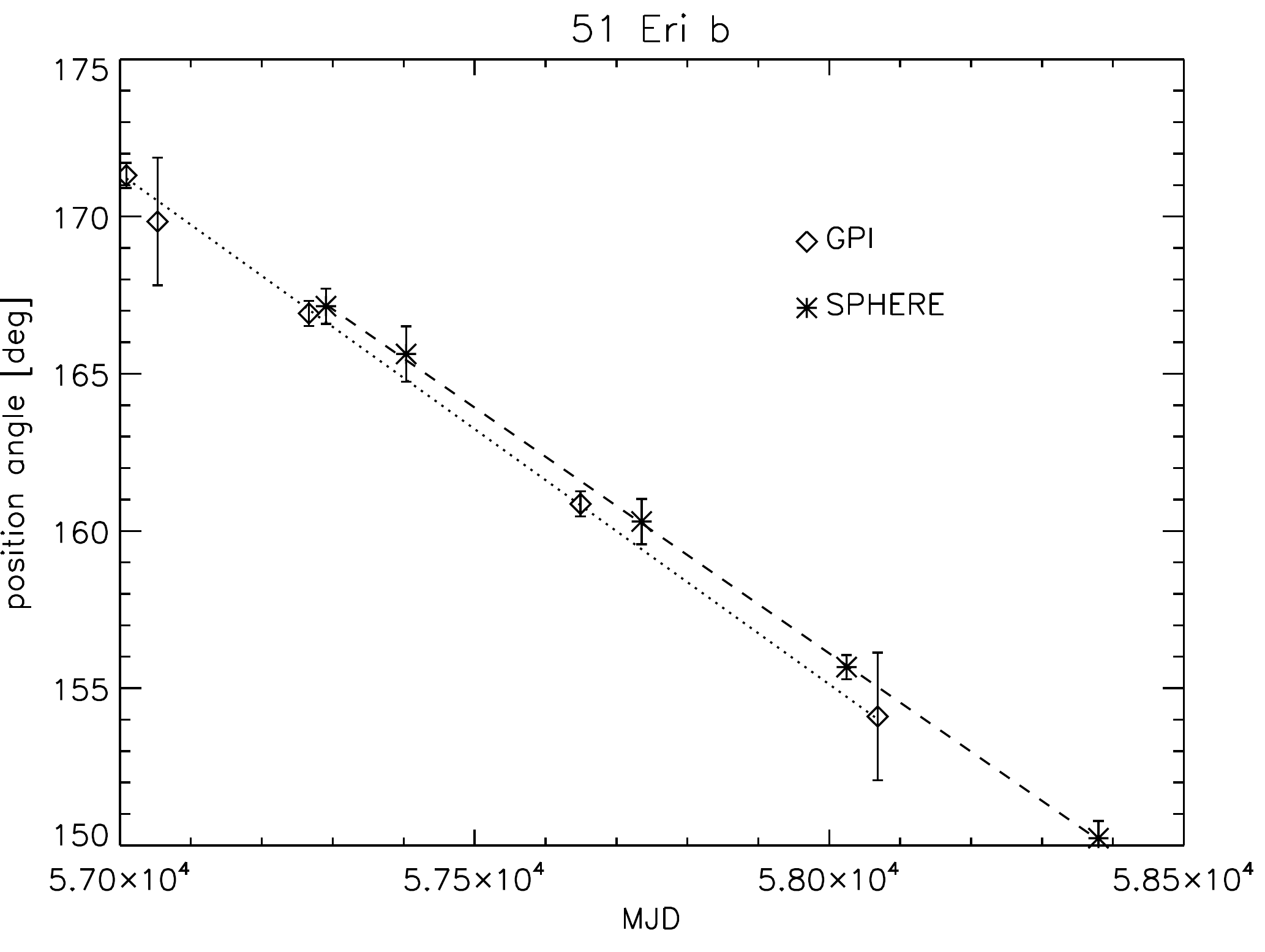}
\caption{{Temporal evolution of the position angle of 51~Eridani~b measured in the SPHERE data (stars) and our analysis of GPI archival data (diamonds). Linear fits are shown for each data series separately (SPHERE: dashed line, GPI: dotted line).}}
\label{fig:51eri_pa_spheregpi}
\end{figure}

{In order to further analyze potential position angle systematics between the SPHERE and GPI data, we reduced all the GPI H-band data of 51~Eridani available in the Gemini archive (eight datasets taken from December 2014 to November 2017) using the GPI data reduction pipeline v1.4.0 \citep{Perrin2014, Perrin2016}, which applies an automatic correction for the north offset of $-$1.00$\pm$0.03$^{\circ}$ measured by \citet{Konopacky2014}. We then post-processed them using SDI+ANDROMEDA. The SDI step was necessary to enhance the S/N of the planet. For this, we selected spectral channels where the planet is not expected to show large fluxes due to strong methane absorption \citep{Macintosh2015, Samland2017, Rajan2017}. We were able to recover the planet in five of the datasets. The GPI position angle measurements shown in Fig.~\ref{fig:51eri_pa_spheregpi} display a negative slope ($-$5.9$\pm$0.3$^{\circ}$) in agreement with the slope measured with the SPHERE data but are offset by 0.65$\pm$0.17$^{\circ}$ toward smaller values. We also note that our GPI measurements in common with \citet{DeRosa2015} are offset by $\sim$0.35$\pm$0.05$^{\circ}$} toward larger values.

{From these analyses, we applied a recalibration in position angle of 1.0$\pm$0.2$^{\circ}$ to the GPI measurements in \citet{DeRosa2015} before fitting the SPHERE and GPI data. Due to the location of 51~Eridani~b, the offset in position angle produces an offset mainly in relative right ascension as seen in the top-left panel of Figure~\ref{fig:51eri_seppa_time}.}

{The large GPI/SPHERE position angle offset that we found in our analysis is likely related to differences in the astrometric calibration of the instruments. It is currently unclear if this position angle offset should be considered systematically when combining SPHERE and GPI astrometry in orbital fits because it is not seen for other targets observed with both instruments and with published observations close in time \citep[HD~95086, HR~2562, $\beta$~Pictoris, HR~8799;][]{Rameau2016, Konopacky2016b, JWang2016, JWang2018, Chauvin2018, Maire2018, Lagrange2019, Zurlo2016}. Further analysis is needed to conclude on this point but is considered to be beyond the scope of this paper.}

\section{LSMC orbital fitting}
\label{sec:orbfamiliesppties}

We drew {1\,000\,000} random realizations of the astrometric measurements assuming Gaussian distributions around the nominal values. We then fit the six Campbell elements simultaneously using the downhill simplex \texttt{AMOEBA} procedure provided in the \texttt{EXOFAST} library \citep{Eastman2013}: orbital period $P$, eccentricity $e$, inclination $i$, longitude of node $\Omega$, argument of periastron passage $\omega$, and time at periastron passage $T_0$. Initial guesses of the parameters were {computed} assuming uniform distributions in log\,$P$, $e$, cos\,$i$, $\Omega$, $\omega$, and $T_0$. We considered no restricted ranges except for the period ($P$=10--1000\,yr). We also included the correction for the bias on the eccentricity and time at periastron passage due to the small orbital arc covered by the data following the method in \citet{Konopacky2016a} {($\sim$34\% of fitted orbits are rejected} when applying this method, because of long periods, large eccentricities, and/or times at periastron passage close to the epochs of the data).

{For the corner plot and the 68\% intervals of the parameters shown in Sect.~\ref{sec:orbit}, we retained for the analysis} all the derived solutions with $\chi_{\rm{red}}^2$\,<\,2. The longitude of node and the argument of periastron passage are restrained in the [0;180$^{\circ}$] range to account for the ambiguity on the longitude of node inherent to the fitting of imaging data alone.

{We note} two broad peak features around $\sim$10$^{\circ}$ and 130$^{\circ}$ in the distribution of the longitude of node, which are associated with eccentricities of $\sim$0.2--0.5 and show correlations with the time at periastron passage. The peak around $\sim$10$^{\circ}$ appears to produce more orbits with $T_0$ around $\sim$\,2005, whereas the peak around $\sim$130$^{\circ}$ appears to produce more orbits with $T_0$ around $\sim$\,2025.

\section{MCMC orbital fitting}
\label{sec:mcmcorbit}

Here we provide the parameter distribution obtained using an MCMC approach \citep[see details in][]{Chauvin2012}. We assumed uniform priors in log\,$P$, $e$, cos\,$i$, $\Omega + \omega$, $\omega - \Omega$, and $T_0$. Ten chains of orbital solutions were conducted in parallel, and we used the Gelman–Rubin statistics as convergence criterion \citep[see the details in][]{Ford2006}. We picked a random sample of 500\,000 orbits from those chains following the convergence.

\begin{figure*}[t]
\centering
\includegraphics[width=.99\textwidth]{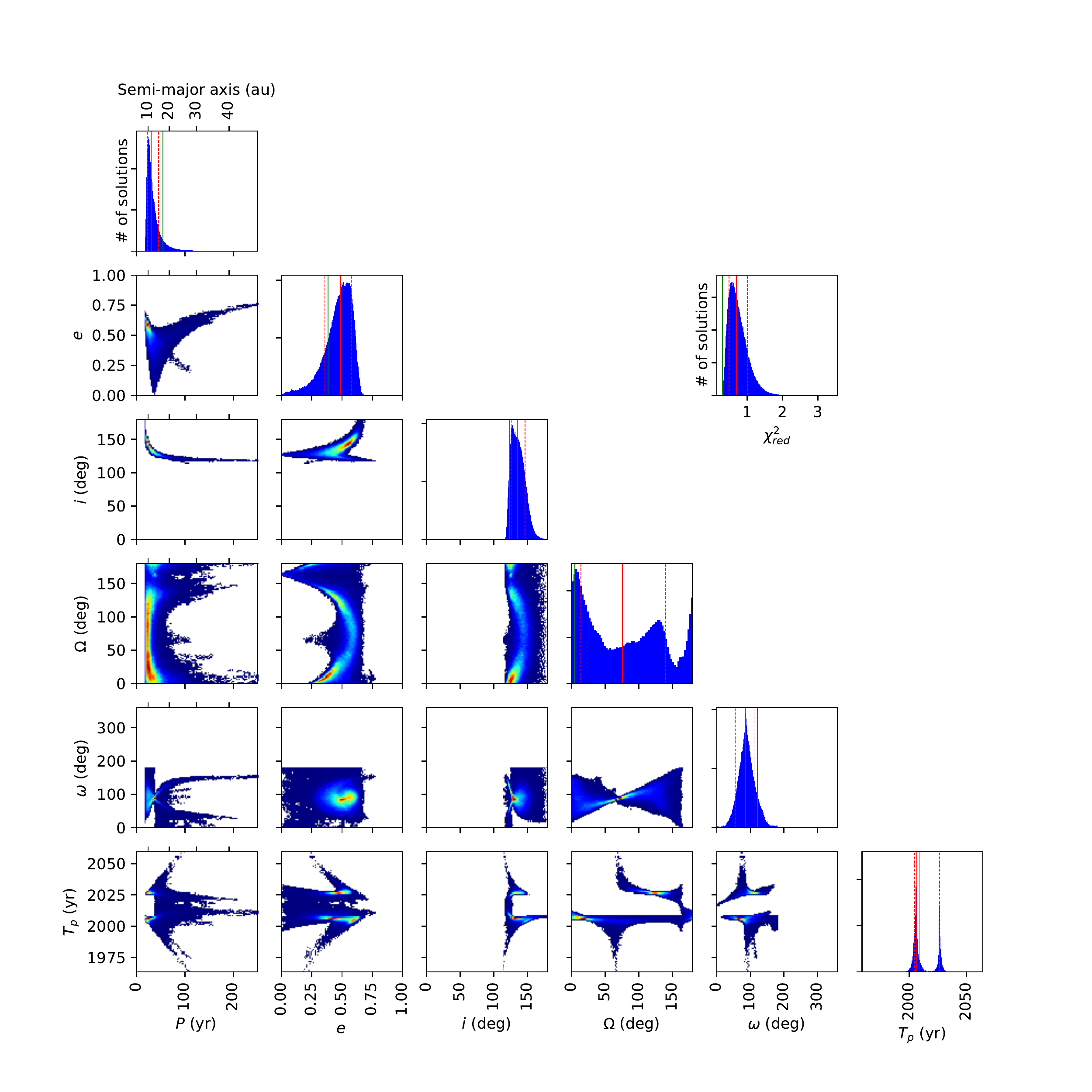}
\caption{{MCMC distributions of the six Campbell orbital elements. The diagrams displayed on the diagonal from top left to bottom right represent the 1D histogram distributions for the individual elements. The off-diagonal diagrams show the correlations between pairs of orbital elements. The linear color-scale in the correlation plots accounts for the relative local density of orbital solutions. The diagram in the top-right part shows the histogram distribution of the reduced $\chi^2$.}}
\label{fig:51eri_cornerplotmcmc}
\end{figure*}

\begin{table}[h]
\caption{{Orbital parameters of 51~Eridani~b derived from the MCMC analysis}.}
\label{tab:orbparamsmcmc}
\begin{center}
\begin{tabular}{l c c c c c}
\hline\hline
Parameter & Unit & Median & Lower & Upper & $\chi^2_{\rm{min}}$ \\
\hline
$P$ & yr & 30 & 23 & 46 & 53 \\
$a$ & au & 12 & 10 & 15 & 17 \\
$e$ & & 0.49 & 0.36 & 0.58 & 0.39 \\
$i$ & $^{\circ}$ & 135 & 126 & 146 & 124 \\
$\Omega$ & $^{\circ}$ & 76 & 14 & 139 & 5 \\
$\omega$ & $^{\circ}$ & 85 & 54 & 111 & 121 \\
$T_0$ & & 2007 & 2005 & 2027 & 2009 \\
\hline
\end{tabular}
\end{center}
\end{table}

\section{OFTI orbital fitting}
\label{sec:oftiorbit}

Here we provide the parameter distribution obtained using a custom {Interactive Data Language (IDL)} implementation of the Orbits For The Impatient (OFTI) approach described in \citet{Blunt2017}. Briefly, we drew random orbits from uniform distributions in $e$, cos\,$i$, $\omega$, and $T_0$ and adjusted their semi-major axis and longitude of node by scaling and rotating the orbits to match one of the measured astrometric points. {As explained in \citet{Blunt2017}, the scale-and-rotate method to adjust the semi-major axis and longitude of node imposes uniform priors in log\,$P$ and $\Omega$.} Subsequently, the $\chi^2$ probability of each orbit was computed assuming uncorrelated Gaussian errors before performing the rejection sampling test. To speed up the procedure in order to obtain a meaningful number of orbits ({29\,870}), we applied the procedure at each iteration over 8\,000 trial orbits simultaneously and we also restrained the prior ranges using the statistics of the first 100 accepted orbits. 

{We cross-checked our code with the OFTI procedure available as part of the Python \texttt{orbitize} package \citep{Blunt2019}}. For verification, we also performed a fit using only GPI data points (2014 December 18, 2015 January 30, and 2015 September 1) from \citet{DeRosa2015} and assuming the same priors for the parameters; we found parameter distributions and intervals very similar to those obtained by these latter authors.

\begin{figure*}[t]
\centering
\includegraphics[width=.99\textwidth]{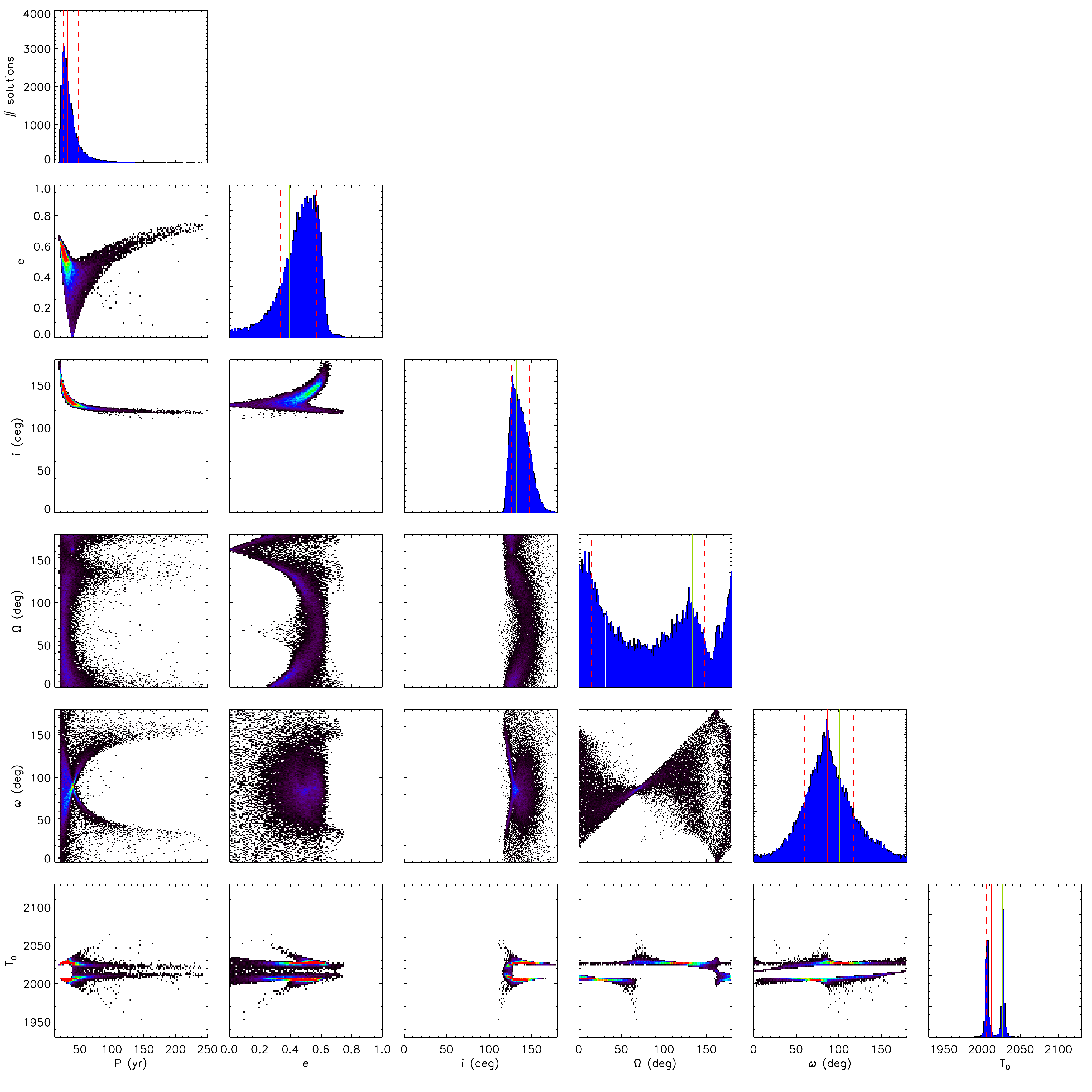}
\caption{{As in Fig.~\ref{fig:51eri_cornerplot} but obtained using the OFTI approach (see text).}}
\label{fig:51eri_cornerplotofti}
\end{figure*}

\begin{table}[h]
\caption{{Orbital parameters of 51~Eridani~b derived from the OFTI analysis.}}
\label{tab:orbparamsofti}
\begin{center}
\begin{tabular}{l c c c c c}
\hline\hline
Parameter & Unit & Median & Lower & Upper & $\chi^2_{\rm{min}}$ \\
\hline
$P$ & yr & 31 & 23 & 47 & 34 \\
$a$ & au & 12 & 10 & 16 & 13 \\
$e$ & & 0.47 & 0.33 & 0.57 & 0.39 \\
$i$ & $^{\circ}$ & 135 & 126 & 147 & 132 \\
$\Omega$ & $^{\circ}$ & 82 & 15 & 148 & 134 \\
$\omega$ & $^{\circ}$ & 86 & 59 & 118 & 101 \\
$T_0$ & & 2012 & 2006 & 2027 & 2026 \\
\hline
\end{tabular}
\end{center}
\end{table}

\newpage

\section{{Stellar rotation axis}}
\label{sec:photometry}

{\citet{Koen2002} estimated a rotation period of 0.65~d for 51~Eridani from \textit{Hipparcos} photometric data without giving an uncertainty. This rotation period was used by \citet{Feigelson2006} with a stellar projected rotational velocity of $v$\,sin\,$i_{\star}$=\,71.8\,$\pm$\,3.6~km\,s$^{-1}$ \citep{Reiners2003} and a stellar radius of 1.5~$R_{\odot}$ to estimate an inclination of 45$^{\circ}$ for the rotation axis of the star.}

\begin{figure*}[t]
\centering
\includegraphics[width=.59\textwidth, angle=90]{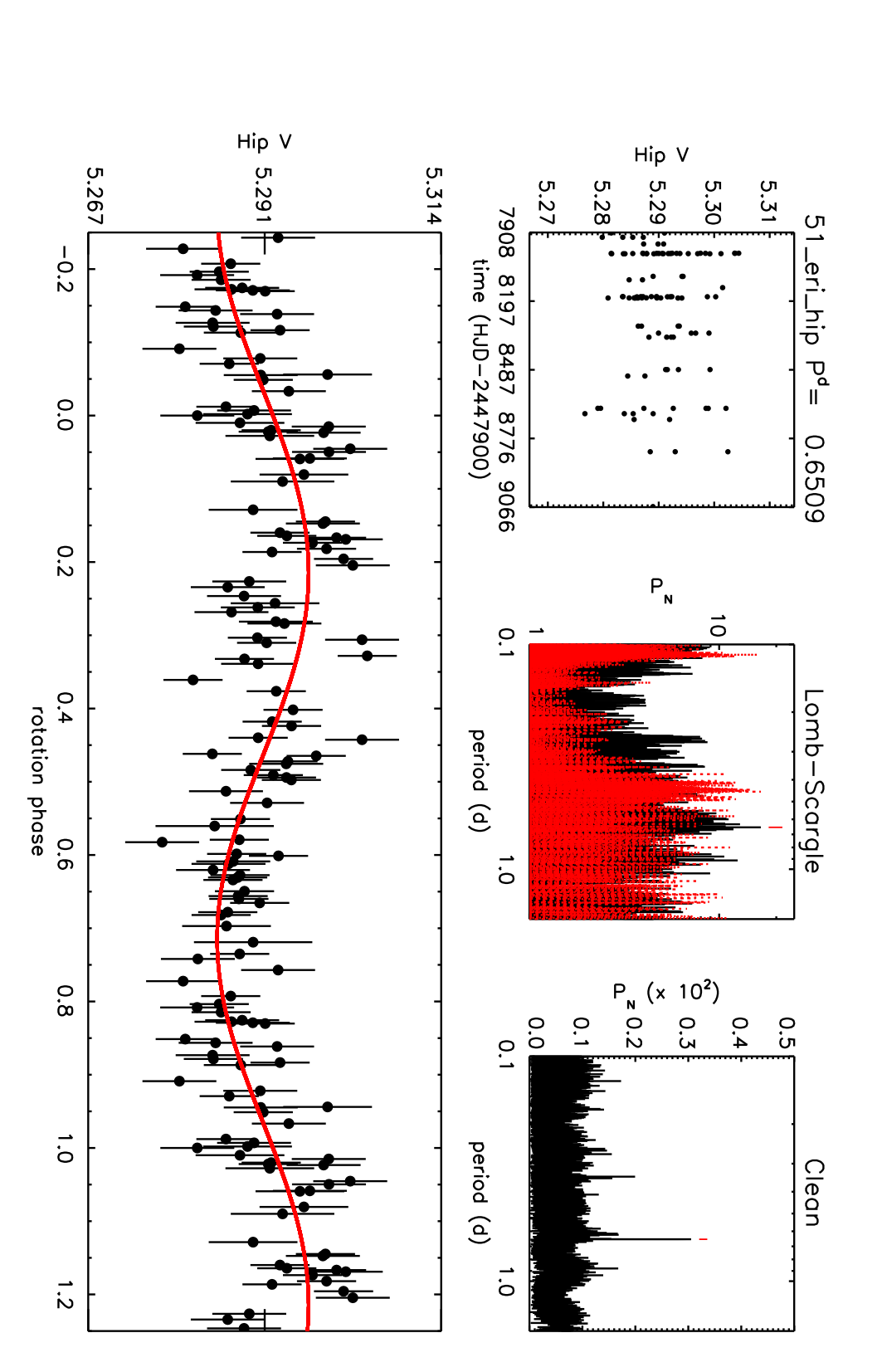}
\caption{{Photometric analysis of 51~Eridani based on \textit{Hipparcos} data. \textit{Top row from left to right}: $V$-band magnitude vs. Heliocentric Julian Day, Lomb-Scargle periodogram, and \textit{CLEAN} periodogram. For the Lomb-Scargle periodogram, we show the spectral window function (in red) and the peak corresponding to the rotation period (red vertical mark). \textit{Bottom panel}: Light curve phased with the rotation period. The solid red curve represents the sinusoidal fit.}}
\label{fig:51eri_hip}
\end{figure*}

\begin{figure*}[t]
\centering
\includegraphics[width=.59\textwidth, angle=90]{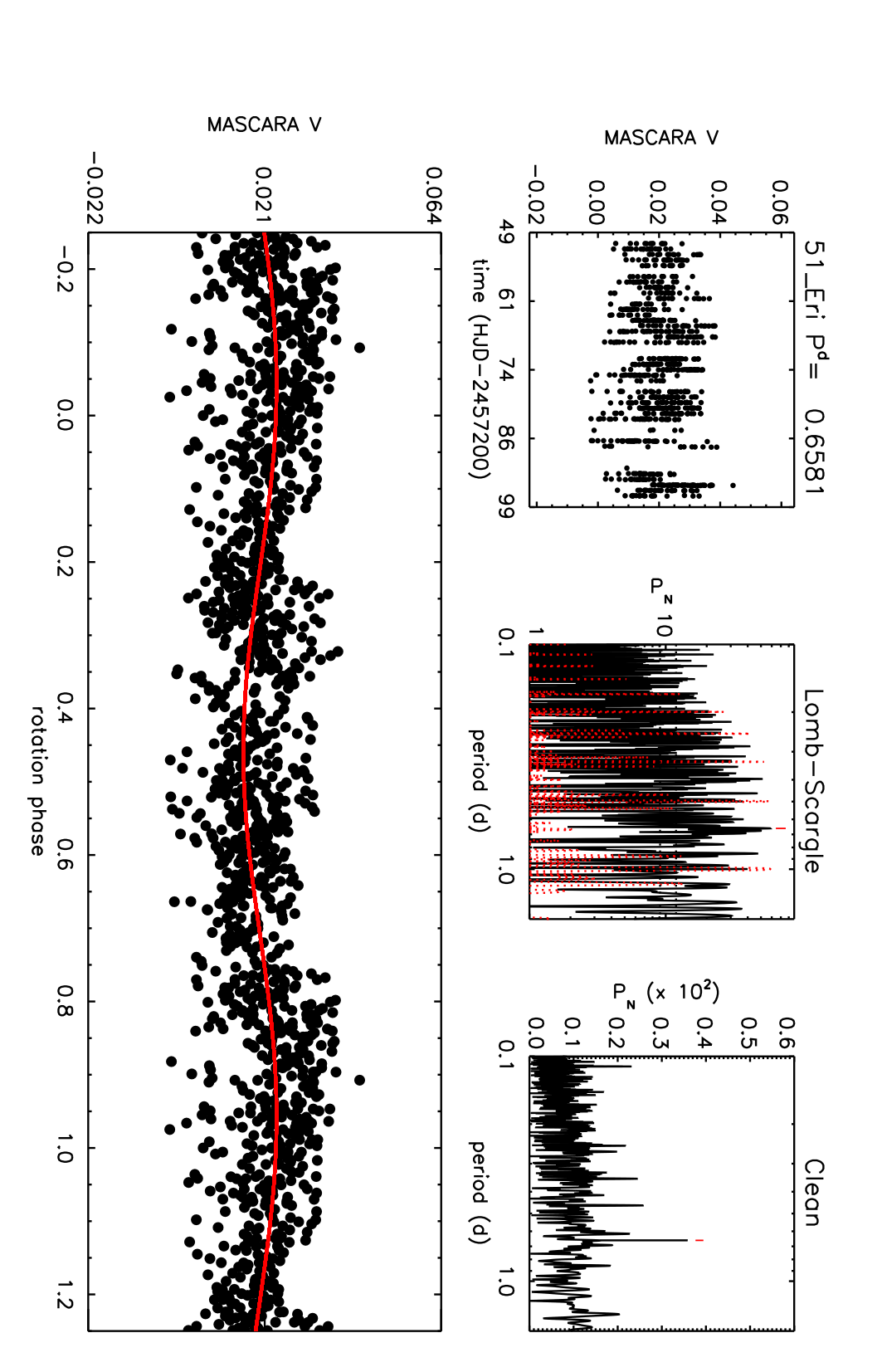}
\caption{{As in Fig.~\ref{fig:51eri_hip} but for the MASCARA data.}}
\label{fig:51eri_mascara}
\end{figure*}

{In order to better constrain the stellar rotation axis and estimate in particular an uncertainty on this parameter, we reanalyzed the \textit{Hipparcos} photometric data \citep{Perryman1997, vanLeeuwen1997}. Figure~\ref{fig:51eri_hip} shows the results. Both the Lomb-Scargle periodogram \citep{Scargle1982} and the \textit{CLEAN} periodogram \citep{Roberts1987} show a peak at $P_{\star}$\,=\,0.65\,$\pm$\,0.03~d. We also analyzed as a cross-check analysis archival data from {the Multi-site All-Sky CAmeRA \citep[MASCARA;][]{Talens2017}} and found a rotation period in good agreement with the value derived from the \textit{Hipparcos} data ($P_{\star}$\,=\,0.66~d, Fig.~\ref{fig:51eri_mascara}). We considered only the \textit{Hipparcos} results in the remainder of the analysis.}

{Using a $V$ magnitude $V$\,=\,5.20~mag, a distance $d$\,=\,29.78\,$\pm$\,0.15~pc, a bolometric correction $BC_V$\,=\,0~mag, and an average effective temperature from the literature $T_{\star}$\,=\,7250~K, we infer a stellar radius $R_{\star}$\,=\,1.53\,$\pm$\,0.04~$R_{\odot}$. Combining the rotation period, the stellar radius, and an average projected rotational velocity $v$\,sin\,$i_{\star}$\,=\,83\,$\pm$\,3~km\,s$^{-1}$ \citep[estimated from an average of the measurements in][84~km\,s$^{-1}$ and 81.2~km\,s$^{-1}$, respectively]{Royer2007, Luck2017}, we infer an inclination of the stellar rotation axis with respect to the line of sight 39$^{\circ}$\,$<$\,$i_{\star}$\,$<$\,51$^{\circ}$ or 129$^{\circ}$\,$<$\,$i_{\star}$\,$<$\,141$^{\circ}$. Given the derived orbital inclination of the planet (126--147$^{\circ}$), this suggests alignment for the spin-orbit of the star-planet system. Using the stellar radius measurement of 1.63\,$\pm$\,0.03~$R_{\odot}$ in \citet{Simon2011} gives 36$^{\circ}$\,$<$\,$i_{\star}$\,$<$\,46$^{\circ}$ or 134$^{\circ}$\,$<$\,$i_{\star}$\,$<$\,144$^{\circ}$, which also suggests spin-orbit alignment}.

\end{appendix}

\end{document}